\documentclass[nolinenumbers,trackchanges]{aastex701}
\usepackage{amsmath}
\usepackage{longtable}
\newcommand{\kms}{km\,s$^{-1}$}
\newcommand{\ms}{m\,s$^{-1}$}
\newcommand{\me}{\mathrm{e}}
\newcommand{\dif}{\mathrm{d}}

\begin{document}
\nolinenumbers

\title{Polarization Observations of a Sample of 6.7\,GHz Methanol Masers }

\author[0000-0002-4846-1741,gname=Paul, sname=Fallon]{Paul Fallon}
\affiliation{Centre for Space Research, North-West University, Private Bag X1290, 
Potchefstroom 2520, South Africa}
\affiliation{UNISA Centre for Astrophysics and Space Sciences (UCASS), College 
of Science, Engineering and Technology, University of South Africa, Florida 
1709, South Africa}
\email[show]{paulfallon@telkomsa.net}

\author[0000-0003-3593-9707,gname=Derck, sname=Smits]{Derck P. Smits}
\affiliation{UNISA Centre for Astrophysics and Space Sciences (UCASS), College 
of Science, Engineering and Technology, University of South Africa, Florida 
1709, South Africa}
\email[show]{derck.smits@gmail.com}

\shortauthors{Fallon \& Smits}


\begin{abstract}
Spectra of 6.7\,GHz methanol masers from 21 pointings of known star-forming 
regions are reported. The $C$-band observations, using the Green Bank 
Telescope in full Stokes mode, have measured how polarization properties 
vary across the maser profiles in each spectrum and vary between different 
epochs of observation. Two-thirds of the sources are observed to have 
6.7\,GHz methanol masers, including one new detection (G240.316+0.071). 
Linear polarization is in the range 0 to 15\% and circular polarization 0 to 
$\pm9\%$, in line with previously reported values. 
The only instances where polarization is not observed is when 
these polarization ranges are below $3\sigma$ detection limits. 
Zeeman splitting is observed in several sources, with 
splitting values derived from velocity separation between RCP and LCP 
components via Gaussian fitting. These values are seen to change with time 
and appear to correspond to changes in the linear and circular polarization. 
The polarization properties varying across the spectra and changing with time 
are most likely due to variations in the magnetic fields.
\end{abstract}


\section{Introduction} 
The methanol (CH$_3$OH) $5_1 - 6_0$ A$^+$ transition produces the 6.7\,GHz 
line, which is observed as maser emission in over 1\,000 massive young 
star-forming regions. Methanol is a diamagnetic molecule and is expected 
to display polarization characteristics in the presence of magnetic 
fields, but because it is non-paramagnetic, both the linear and circular 
polarization fractions are small. Based on the calculations of \citet{LVS18} 
the internal rotation of the molecule produces a complicated hyperfine 
structure that leads to a large range in values of the Landé $g$ factors 
for the hyperfine components. It is not certain which of the eight hyperfine 
transitions dominates the 6.7\, GHz maser, though \cite{LVS18} assumed the 
$F=3 \rightarrow 4$ component would be favored. \cite{SVL22} use this in 
determining their Zeeman coefficient and the magnetic field estimates, but 
note that as this hyperfine transition has the largest Landé $g$ factor, 
as their magnetic field values are a lower limit.

Simulations by \cite{DVLS20} show that both linear and circular 
polarization fractions depend on the hyperfine component and the degree 
to which it is being pumped. They compare their modeled results with 
observed high linear and circular polarization fractions and suggest 
that other hyperfine transitions may be dominant. By comparing Zeeman 
splitting from nearby excited OH (exOH) and methanol masers \cite{KSB25} 
attempt to determine the dominant 6.7\,GHz hyperfine transition, though 
this is based on an assumption that both masers probe the same magnetic 
field. 

The first report of polarization in methanol 6.7\,GHz masers was made by 
\citet{E02} who used the Australia Telescope Compact Array (ATCA) to 
detect linear polarization of $P_\textrm{l}<10$\% in four sources. Since then 
there have been numerous polarization observations. \citet{DM12} mapped 
Stokes $I,\,Q$ and $U$ signals of 10 methanol maser sources with the ATCA. 
Linear polarization percentages in their sample range from a few to 
$\sim10\%$. \citet{GCM15} found 60 sites of methanol masers with the ATCA, 
of which 22 showed linear polarization. No circular polarization was found 
in any of their sources. The source \object{G339.88--1.26} was observed with 
the Long Baseline Array (LBA) by \citet{Dod08} to create $I,\ Q\ \mbox{and}\ 
U$ maps from which linear polarization of up to 10\% was found. 

The MultiElement Radio-Linked Interferometer Network (MERLIN) was used by 
\citet{VHC06} to map the polarization properties of masers in W3(OH). Linear 
polarization $P_\textrm{l} < 8\%$ was found, with an error-weighted median 
value $\langle P_\textrm{l} \rangle = 1.8(7)\%$, while circular polarization, 
although not detected, was determined to have an upper limit $P_\textrm{c} 
< 2\%$.  The ultracompact H\,II region \object{ON1} was observed with MERLIN 
by \citet{GRV07}, who found it displayed linear polarization $P_\textrm{l} 
\sim1\%$ in two of four spots\footnote{We have used the term "maser 
spot" to refer to an observed single spacial point of maser emission. The 
term "maser feature" is often used rather than our description as the 
observed point of emission may comprise multiple  regions of unresolved
maser activity.}, but with position angles (PAs) differing by 
90\degr. A tentative first detection of circular polarization with 
$P_\textrm{c} = 0.65\%$ produced a Zeeman splitting of $\Delta V_\textrm{Z} = 
0.9(3)$\,\ms. \citet{HSD08} observed \object{DR21(OH)} and \object{DR21(OH)N} 
in full Stokes mode with MERLIN. Linear polarization $P_\textrm{l} = 4.9\%$ 
and 4.5\% with a PA of --40\degr\ was found in the two spots in DR21(OH)N 
while DR21(OH) was unpolarized. 

The Effelsberg 100m telescope was used by \citet{V08} to measure the 
Zeeman splitting in a sample of 24 bright northern hemisphere masers. 
Significant Zeeman splitting was found in 17 sources, and was determined 
using a cross-correlation method described by \citet{MMKG05} applied 
to the entire spectrum or parts thereof. This produced a flux-weighted 
average value for the splitting, but does not require that Gaussians be 
fitted to the individual peaks in the spectra.
Variability in the Zeeman splitting of the episodically flaring methanol 
maser in the source \object{G09.62+0.20} was measured by \citet{VGG09} using 
the Effelsberg 100m telescope over a fortnight surrounding a maximum in the 
flaring activity. In contrast, two other maser features in this object 
showed constant Zeeman splitting, suggesting a steady $B_{\parallel}$ in 
these regions. 

The European VLBI network (EVN) was used to measure the linear polarization 
and Zeeman splitting in the star-forming region \object{W75N} by 
\citet{SVD09}. They detected 10 methanol maser spots near the source labeled 
VLA 1. The linear polarization, which had fractional polarization $P_\textrm{l} 
\sim 2 - 5\%$ in their group A and $P_\textrm{l} \sim 1-2\%$ in groups B and C, 
aligned with the large-scale molecular outflows in the source. Zeeman splitting 
with $\Delta V_\textrm{Z} = 0.80(3), 0.75(13), 0.53(4)$\,\ms\ was measured 
using a cross-correlation method in three maser features.

\citet{VST10} used MERLIN to look at the masers in \object{Cepheus A HW2}.
They deduced the three-dimensional structure and strength of the magnetic
field from their full Stokes measurements. They found a magnetic field 
$B\sim23$\,mG aligned along the protostellar outflow and perpendicular to 
the molecular and dust disk. 

\citet{SVT11} used the EVN and MERLIN to map \object{NGC 7538 IRS1}. Of 
49 masers detected, 20 had linear polarization, and Zeeman splitting was 
found in 3 of the spots.

A set of five observing sessions using the EVN in full Stokes mode were 
undertaken by \citet{SVL12, SVL13, SVL15, SVL19, SVL22}. A statistical study 
of all 31 sources shows that the average linear polarization of the 233 spots 
lies in the range $\langle P_\textrm{l} \rangle = [1.0,\ 2.5]\%$, while the 
circular polarization based on 33 spots lies in the range $\langle P_\textrm{v} 
\rangle = [0.5,\ 0.75]\% $. They found Zeeman splitting typically in the range 
0.5 \ms\ to 2.0 \ms\ and estimate corresponding lower limit values for the 
magnetic field. 

The source IRAS 18089--1732 was observed with MERLIN using full Stokes mode 
in 2008 March and April and two circular modes in July by \citet{DVS17}. The 
linear polarization $P_\textrm{l} < 10\%$ in March and April, but showed 
changes in $P_\textrm{l}$ and the PA from one epoch to the next. A tentative 
detection of circular polarization was made in one bright component (labeled 
F.06 in their paper), which they resolved into two Gaussian components that 
appeared to reverse in polarity between the March and April observations. 
They suggest possible explanations for the circular polarization reversal 
which are discussed in Section \ref{sec:pol_changes}. The linear and circular 
polarization observations show that the magnetic field in the 6.7\,GHz methanol 
maser region is consistent with the magnetic field constrained by Submillimeter 
Array (SMA) dust polarization observations.

This paper presents full Stokes observations of a sample of 6.7\,GHz 
methanol masers conducted with the Green Bank Telescope (GBT). We report 
on the circular and linear polarization parameters, how the polarization 
characteristics vary across the spectra, and show changes in flux 
density and polarization between different epochs.

\begin{table*}
	\centering
	\caption{Source list in order of increasing RA containing source 
        name, J(2000) coordinates, velocity, polarization calibrator, 
        observing time, rms noise and the dates observed. }
	\label{tab:source}
	\begin{tabular}{clccccccc} 
		\hline
	No.	&Source   &R.A.(J2000)   &Dec.(J2000)   &Velocity   &Polarization  &\multicolumn{1}{c}{Time}  &RMS   &Date Observed   \\
         &        &(hh mm ss.s)    &$\degr\ \ '\ \ ''$   &(\kms)  &Calibrator &\multicolumn{1}{c}{(minutes)}  &(mJy)   &(20yy mmm dd)      \\
		\hline
	  1	 &G108.758-0.986   &22 58 47.5   &+58 45 01.8   &--46   &3C48  &50  &39  &23 Jul 31 \\
    2a &G111.526+0.803   &23 13 33.1   &+61 29 15.2   &--59   &3C48  &68  &17  &23 Apr 21 \\ 
    2b &                 &             &              &       &      &42  &39  &23 Jun 23 \\
    3  &G111.532+0.759   &23 13 43.9   &+61 26 55.7   &--59   &3C48  &62  &30  &23 Jul 07 \\
    4  &G111.542+0.777   &23 13 45.3   &+61 28 10.0   &--58   &3C48  &60  &38  &23 Aug 03 \\
    5  &G126.715--0.822  &01 23 33.7   &+61 48 49.2   &--12   &3C48  &64  &26  &23 Nov 21 \\
    6  &G133.715+1.215   &02 25 40.6   &+62 05 50.5   &--40   &3C48  &64  &46  &23 Dec 24 \\
    7  &G133.947+1.064   &02 27 03.7   &+61 52 25.0   &--44   &3C48  &68  &33  &23 Jul 28 \\
    8  &G141.918+1.902   &03 27 28.6   &+58 53 48.1   &\ --8  &3C48  &64  &27  &23 Jul 07 \\
    9  &G163.078--1.926  &04 49 46.7   &+41 39 05.5   &--15   &3C138 &64  &26  &23 Jul 06 \\
    10 &IRAS 05137+3919  &05 17 12.8   &+39 22 05.0   &--20   &3C138 &64  &26  &23 Dec 31 \\
    11 &G173.482+2.446   &05 39 13.0   &+35 45 51.0   &--16   &3C138 &64  &12  &23 Apr 23 \\
    12 &G211.567--19.28  &05 39 56.0   &--07 30 27.7  &+2     &3C138 &68  &14  &23 Apr 21 \\
    13 &IRAS 05382+3547  &05 41 37.4   &+35 48 49.0   &--24   &3C138 &68  &12  &23 Apr 24 \\ 
    14a &IRAS 05392--0214 &05 41 44.8  &--02 13 22.7  &+11    &3C138 &68  &12  &23 Apr 22 \\
    14b &                &             &              &       &      &64  &27  &23 Jun 24 \\
    15 &G183.349--0.575  &05 51 11.0   &+25 46 15.9   &\ --6  &3C138 &64  &28  &23 Jul 06 \\
    16a &G213.705--12.60 &06 07 47.7   &--06 23 01.2  &+11    &3C138 &70  &30  &23 Dec 23 \\
    16b &                &             &              &       &      &70  &31  &24 Feb 24\\
    17 &G189.030+0.783   &06 08 36.1   &+21 20 28.0   &\ +3   &3C138 &60  &13  &23 May 30 \\
    18 &G188.946+0.886   &06 08 53.3   &+21 38 29.0   &+11    &3C138 &80  &16  &23 May 29 \\
    19 &G196.454--1.677  &06 14 37.1   &+13 49 36.0   &+15    &3C138 &60  &28  &23 Jul 28 \\
    20 &G232.620+0.996   &07 32 09.8   &--16 58 13.0  &+23    &3C138 &64  &33  &23 Jul 06 \\
   21a &G240.316+0.071   &07 44 53.2   &--24 07 39.0  &+63    &3C138 &80  &24  &23 Oct 25 \\
   21b &                 &             &              &       &      &88  &25  &24 Jan 27\\    
		\hline
	\end{tabular}
\end{table*}

\section{Observations}
We observed 21 pointings with Right Ascensions (RA) in the range from 23$^h$ 
to 08$^h$ to meet the LST range for filler sources specified by a GBT 
Special Proposal Call. Table~\ref{tab:source} lists the objects and their 
J2000 coordinates, the central velocity of the spectra, the linear 
polarization calibrator used, the time on source, the rms noise achieved 
and the date(s) when the observations were done. Most sources were only 
observed once, but where observing opportunities allowed, sources were 
observed twice to see if flux density, central velocity or polarization 
properties changed in the months between observations. The source 
\object{G213.705--12.60} (\object{Mon R2 IRS3}) was also observed as part 
of this programme initially, but when it started flaring at 4.765\,GHz 
it was observed multiple times. The time series results of these data will 
be reported separately.

All observations reported here were made with the 100m GBT using the 
$C$-band receiver with the Versatile GBT Astronomical Spectrometer (VEGAS) 
backend in full Stokes mode. At 6.7\,GHz the FWHM beam size is 
approximately 1.9 arcminutes. The VEGAS backend contains eight bands each 
of which was operated in mode 15, providing a bandwidth of 11.72\,MHz over 
32\,768 channels, corresponding to a frequency resolution of 357.7\,Hz. 
One frequency band was used for the methanol maser line, centered at the 
methanol rest frequency of 6.668\,518\,GHz. This gives a velocity 
resolution of 0.0161\,\kms\ and in frequency-switching mode the velocity 
coverage is 527\,\kms. The other seven bands in the VEGAS backend were 
used to observe the exOH lines of the $^2\Pi_{1/2}\ J = 1/2$ level at 
4.765, 4.750 and 4.660\,GHz and the $^2\Pi_{3/2}\ J = 3/2$ level at 6.016, 
6.031, 6.035 and 6.049\,GHz. The results of these measurements will be 
reported separately (Smits \& Fallon, in preparation). 

The Stokes spectra are obtained using orthogonal linear feeds and calibrated 
using a Mueller matrix as described in \citet{FSG23}. A standard linear 
polarization calibrator, either 3C48 or 3C138, was observed as part of each 
observing session to confirm and refine the calibration. The calibration 
process enables linear polarization fraction accuracy of $\sim 0.5\%$ and 
position angle accuracy of $\sim1$\degr\ for the reference calibrators with 
intensity of 3 to 6 Jy \citep{FSG23}. In addition to the Mueller matrix 
calibration, further refinements of the Stokes $V$, described by \citet{SF25}, 
are done to ensure that leakage of Stokes $I$ into Stokes $V$ is corrected. 
This results in a circular polarization fraction calibration accuracy better 
than $0.5\%$.

The spectra are recorded every 2 minutes. Analysis involves shifting and 
folding each 2 minute scan, an initial intensity calibration and background 
subtraction using a third order polynomial. The Mueller matrix calibration 
is applied to each scan which are then added to produce the on-sky $I$, $Q$, 
$U$ and $V$ spectra. 

Many of our sources have been monitored at 6.7\,GHz by 
Ibaraki\footnote{http://vlbi.sci.ibaraki.ac.jp/iMet/} which will 
henceforth be referred to as iMet. We note our flux densities are a factor 
$\sim1.6$ below values published by iMet for two sources on similar 
observation dates. Comparison to known reference calibrators confirm 
the accuracy of our values, and communication with iMet authenticates their 
calibration, so the source of the difference is unclear.

\section{Results} \label{sec:results}
All the Stokes $I$ spectra were fitted with a varying number of Gaussian 
profiles (using the least squares routine in GBTIDL) in an attempt to 
reduce the residuals to the order of the rms noise level. Because this 
table is long and not relevant to the discussion, the fitted parameters 
are listed in the Appendix. When residuals are larger than the noise, this 
is usually an indication that there are more spots of emission than the 
number of Gaussians used for the fitting. For methanol maser spectra, 
interferometric observations have confirmed that there are often many more 
spots of maser emission than are apparent as peaks in single-dish spectra. 

Linear and circular polarization, shown for each source in Figures 
\ref{fig:G108_7} -- \ref{fig:G240_7}, are determined from the Stokes spectra 
in the standard manner\footnote{Note a different circular polarization 
convention has been used by several other authors when an S-shaped 
Stokes $V$ spectrum is observed. Their circular polarization definition 
and the one we use here are not comparable -- see Appendix \ref{App:CirPol}.}
\citep{RH21} using
\begin{align}
\text{linear polarization fraction} &= P_\textrm{l} =\frac{\sqrt{Q^2 + U^2}}{I}, 
      \text{ with} \\
\text{position angle} &= \text{PA} =\frac{1}{2} \tan^{-1}\left(\frac{U}{Q}\right), 
      \text{ and} \\
\text{circular polarization fraction} 
&= P_\textrm{c} =\frac{V}{I} \label{eq:CirPol}.
\end{align}

When there is no detectable polarization towards a source, only the Stokes 
$I$ component has been plotted. The plots in Figures 1 -- 14 use a black solid 
line for the original data and a red solid line for the fitted spectra. Rather 
than showing the Stokes $Q$ and $U$ components for linear polarization, we 
have converted the parameters into percentage (\%) polarization (magenta) and 
PA (cyan). Stokes $V$ and percentage circular polarization ($P_\textrm{c}$) 
are plotted in dark green. PA, $P_\textrm{l}$ and $P_\textrm{c}$ values are 
plotted when the Stokes $Q$, Stokes $U$ or Stokes $V$ spectrum is above a 
$3\sigma$ detection level, and the rms noise is below a polarization fraction 
of 0.5\%. When RCP and LCP spectra are presented they appear as blue and
dashed magenta respectively. 

The circular components of the signal are shown in the figures and 
determined using RCP = $(I + V)/2$ and LCP = $(I - V)/2$. The Gaussian 
profiles fitted to Stokes $I$, but with half the height, were used as 
starting parameters for fits to RCP and LCP spectra. The final fits to RCP 
and LCP profiles had only small deviations from the initial parameters. 

In many of our spectra, RCP and LCP profiles exhibit intensity differences 
significantly larger than the measurement error of the signal. This produces a 
non-zero Stokes $V$ that is observed in many maser profiles, particularly in 
OH 1.7 and 6\,GHz spectra \citep{D74, S94, CV95, BDWC97, ARM00}. Our Mueller 
matrix calibration process ensures that this is not a calibration error, and 
is confirmed by finding both RCP $>$ LCP and LCP $>$ RCP in the same spectra 
in some instances (see section \ref{sec:w3oh}, for example). 

When RCP and LCP have similar intensity, Zeeman splitting can result in a 
characteristic S-shaped Stokes $V$ spectrum which is proportional to the velocity 
(or frequency) derivative of Stokes $I$ \citep{GSS60, TH82, V08, SF25}. This 
S-shaped profile is generally small and can be obscured by RCP and LCP intensity 
differences, or overlapping maser profiles, which is the case for the all the 
spectra presented in this paper.

For cases where the Gaussian profile is a distinct peak in the maser spectrum, 
and if the central velocities of the RCP and LCP profiles are shifted equally in 
opposite directions from the Stokes $I$ velocity, then it is assumed that the 
separation of the RCP and LCP velocities is the result of Zeeman splitting. 
The magnetic field ($B_{\parallel}$) along the observer's line-of-sight (LOS) 
causes the Zeeman splitting and has been determined from the velocity separation 
as described in section \ref{sec:mag_fields}. 

Determination of Zeeman splitting has also been attempted using the 
cross-correlation method described by \cite{MMKG05}. We have tested 
this method on the spectra of single maser spot described in 
\cite{SF25} and are able to determine reliable values for the Zeeman 
splitting. However, the results are only accurate when the RCP and LCP 
cross-correlation range is $\gtrsim 1.5$ the FWHM either side of the 
maser's central velocity. As our methanol spectra have a complex 
combination of maser lines and the Zeeman splitting can differ 
between velocity adjacent maser features (refer to sections 
\ref{sec:Z_split} and \ref{sec:mag_fields}), use of the 
cross-correlation method for our methanol spectra results in 
inaccurate outcomes when compared to the Gaussian fitting approach 
described above. 

\subsection{Non-detections}
Of our 21 pointings, 6 sources had no detections of 6.7\,GHz emission. The 
sources with non-detections are: \object{G126.715--0.822}, 
\object{G133.715+1.215}, \object{G141.918+1.902}, \object{G163.078--1.926}, 
\object{G211.567--19.28}, and \object{IRAS 05392--0214}. No 6.7\,GHz methanol 
masers have been reported in these sources. G133.715+1.215 has a 4.765\,GHz 
exOH maser, and G141.918+1.902 has 6.031 and 6.035\,GHz exOH masers, which 
will be reported separately. The other sources had no detectable emission 
or absorption at any of our other observed frequencies and will not be 
discussed further.

\subsection{G108.758--0.986 (IRAS 22566+5828)}

\begin{figure*}[ht!] 
\centering
\plottwo{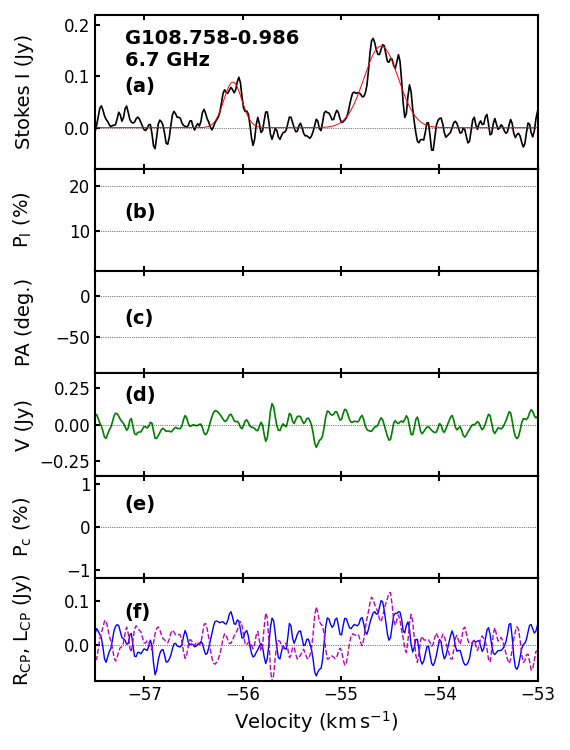}{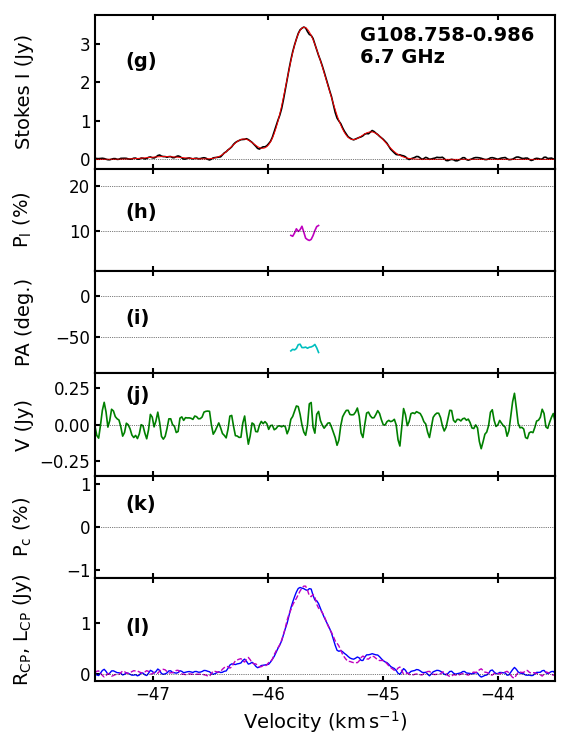}
\caption{Spectra of G108.758--0.986 showing (a) Stokes $I$ (black) with 
fitted Gaussians (red), (b) $P_\textrm{l}$, (c) PA of the linear 
polarization, (d) Stokes $V$, (e) $P_\textrm{c}$, and (f) the RCP 
(blue) and LCP (dashed magenta) components for emission between $v = 
[-57.5,\ -53]$\,\kms, and (g) Stokes $I$, (h) $P_\textrm{l}$, and (i) 
PA of the linear polarization, (j) Stokes $V$, (k) $P_\textrm{c}$, and 
(l) the RCP and LCP components for emission between $v = [-47.5,\ 
-43.5]$\,\kms. Note that top and bottom panels have different scales 
in the left and right hand plots.}
\label{fig:G108_7}
\end{figure*}

The first detection of 6.7\,GHz methanol masers is due to \citet{SHK00} 
who found a 2.8\,Jy source at $v\,=\,-45.7$\,\kms. The catalogue of 6.7\,
GHz masers compiled by \citet{SWB12} shows masers distributed between 
$v\,=\,[-57.2, -44.3]$\,\kms\ with a peak flux density of $\sim3.4$\,Jy 
at $v\,=\,-45.58$\,\kms. The VLA observations of \citet{HMW16} 
identified eight maser spots in a range of velocities between $[-47.5, 
-45.0]$\,\kms\ with a peak flux density of 1.90\,Jy. The iMet data show 
that the masers in this source are highly variable. No polarization 
measurements for this source were found in the literature.

Our spectra for this source are shown in Figure \ref{fig:G108_7}.  There 
is a possible weak maser at $v = -56.1$ ($2.3\sigma$) and another at 
$-54.6$\,\kms\; both are  fitted with a single Gaussian and are weaker 
than in the \citet{SWB12} spectrum. The width and shape of the stronger 
peak at $v = -54.6$\,\kms\ suggest it could be a combination of two spots.

The spectra shown on the right hand side of Figure \ref{fig:G108_7} have 
velocities spread between $[-46.9, -44.9]$\,\kms\ which is different to the 
spectrum of \citet{SWB12}. None of our fitted Gaussians match the velocity 
of spots identified by \citet{HMW16}, however, this is not surprising given 
that this source is considered to be highly variable. On the day of our 
observations, MJD60156, iMet registers the $v = -45.65$\,\kms\ flux density 
as $S_{\nu} = 5$\,Jy. Linear polarization was detected, as can be seen in 
Figure \ref{fig:G108_7}. 

No circular polarization was measured in these spectra to an rms noise 
limit of $\sim50$\,mJy, which equates to 2\% for the 3\,Jy peak at 
$v = -45.7$\,\kms. 

\subsection{NGC 7538 Complex (G111.526+0.803, G111.532+0.759 and G111.542+0.77)}
Methanol masers discovered by \citet{M91} lay between $v = -62$ and $-54$\,
\kms\ and had a peak flux density $S_{\nu} = 346$\,Jy. Maps made using the 
EVN by \citet{MBC98} identified 11 spots of emission which they attributed 
to evidence of a disc. The masers had velocities in the range $[-60.7, -55.5]$\,
\kms, and had a peak flux density $S_{\nu} = 130$\,Jy. \citet{V08} looked at 
the Zeeman splitting in this source with the Effelsberg 100m telescope. The 
peak flux density was $S_{\nu} = 233$\,Jy at $v = -56.1$\,\kms\ and Zeeman 
splittings of $\Delta V_\textrm{Z} = 0.79(3),\, 0.78(4),\, 0.59(2)$\,\ms\ at 
velocities of $v = -56.1,\, -58.0,\, -56.0$\,\kms\ were found. \citet{SVT11} 
used MERLIN and the EVN in full Stokes mode to map masers in the IRS1 region. 
They found 13 spots with MERLIN and 49 with the higher resolution 
of the EVN. 
There were small changes in the shape of their spectra between the 2005 December 
and 2009 November observations using MERLIN and the EVN, respectively. The peak 
flux density was $S_{\nu} \sim 200$\,Jy at a velocity of approximately 
$-56$\,\kms. The VLA observations of \citet{HMW16} identified 43 maser spots 
with a peak intensity of 143\,Jy\,beam$^{-1}$ at $v = -58.13$\,\kms; 12 spots 
lay between $v= -62.0$ and --60.0\,\kms. \citet{ASS23} classified this source 
as moderately variable. 

\begin{figure*}[ht!] 
\centering
\plottwo{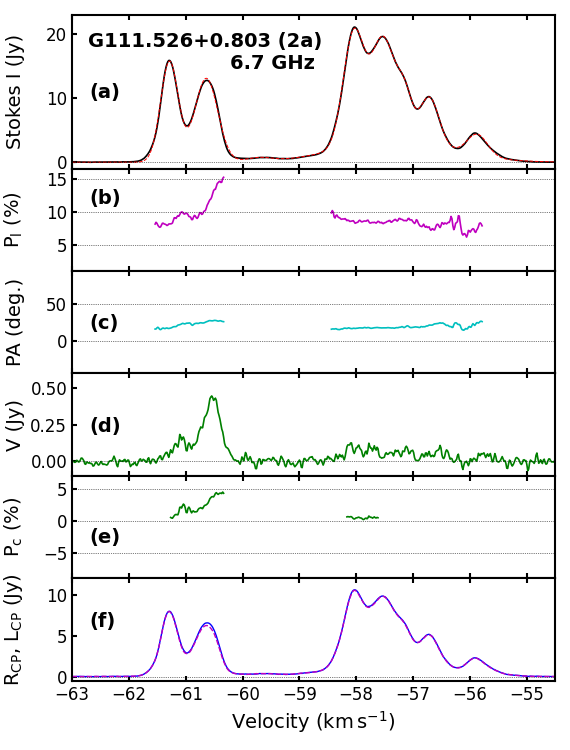}{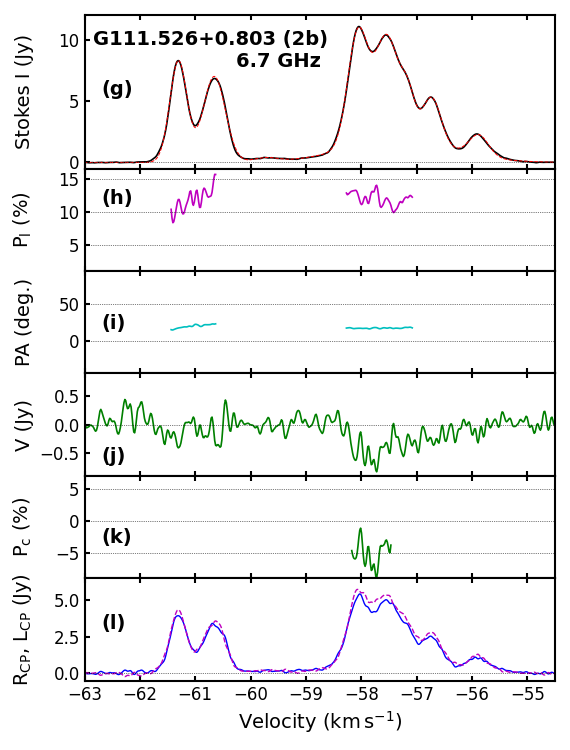}
\caption{Spectra of G111.526+0.803 at the two epochs listed in Table \ref{tab:source} 
showing (a) and (g) Stokes $I$ (black) with fitted Gaussians (red), (b) 
and (h) $P_\textrm{l}$, (c) and (i) PA of the linear polarization, (d) and 
(j) Stokes $V$, (e) and (k) $P_\textrm{c}$, and (f) and (l) the RCP (blue) 
and LCP (dashed magenta) components. }
\label{fig:G111_526}
\end{figure*}

\begin{figure*}[h!] 
\centering
\plottwo{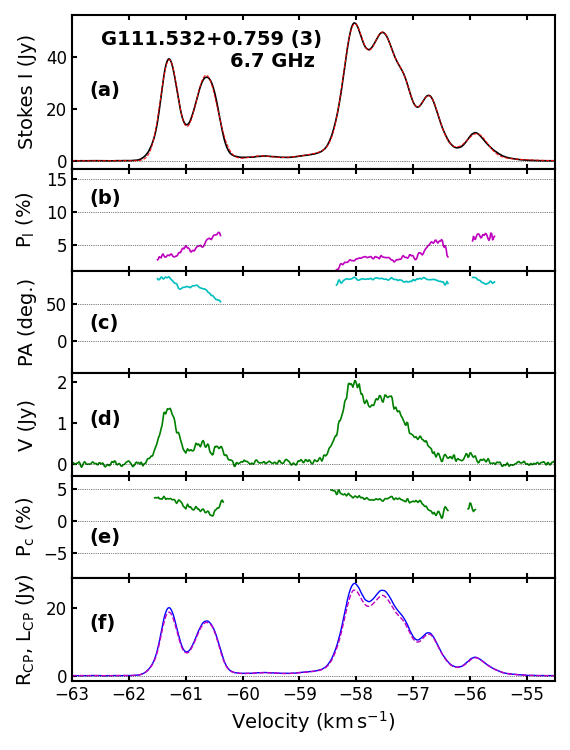}{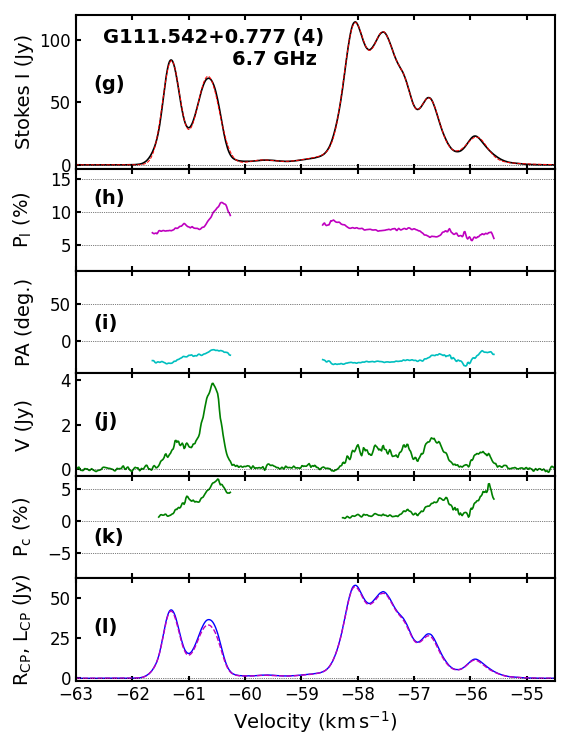}
\caption{Spectra of G111.532+0.759 (left) and G111.542+0.777 (right) 
with the same labelling as in Figure \ref{fig:G111_526}.} 
\label{fig:G111_532}
\end{figure*}

Our observations in Figures \ref{fig:G111_526} and \ref{fig:G111_532} pointed 
at three positions in this complex. According to the map of \cite{OTN04}, our 
pointings are centered on IRS4 (\object{G111.526+0.803}), IRS11 
(\object{G111.532+0.759}) and IRS 1 -- 3 (\object{G111.542+0.777}). Because 
these pointings are so close together, the same masers are seen in all three 
observations. The overall shape of the spectra do not change from one pointing 
to the next, but the flux density differs because the sources appear in different 
parts of the beam and possibly because of time based flux density 
variations. IRS 1 is the brightest object in the cluster and is identified as 
an ultra-compact HII region. IRS 11 is the site of a maser discovered by 
\citet{PMM06}, and is suspected to be a star-forming region younger than 
that in IRS 1. 

In addition to the spectra shown in Figures \ref{fig:G111_526} and 
\ref{fig:G111_532} there is a 1.95\,Jy maser at $v = -53.0$\,\kms\ (see 
Gaussian fits in Table~\ref{tab:GaussFit}) which is polarized but with values 
different from the prominent masers in the plots. There are at least 7 masers 
in the range $[-52.5, -48]$\,\kms\ with flux densities between 0.1 and 0.3 Jy 
but their SNR is too low to provide polarization detail. These features peak 
in the G111.532+0.759 pointing, and are also seen in the iMet monitoring 
spectra. 


The two main peaks with velocities $<-60$\,\kms\ can each be fitted with a 
single Gaussian, and have FWHM of $\sim0.37$ and 0.47\,\kms, which are 
broad profiles for maser lines. The peak velocities of these features vary 
by more than the uncertainties from one epoch to the next. \citet{SVT11} 
found 21 maser features in this velocity range, with FWHM varying over the 
range $\Delta v = [0.12,\ 1.08]$\,\kms. Only one spot, with a velocity $v = 
-60.49$\,\kms\ and a flux density $S_{\nu} = 16.82(3)$\,Jy, had a measurable 
Zeeman splitting of $\Delta V_\textrm{Z} = 2.7(3)$\,\ms. \citet{HMW16} found 
12 features, and although these might have changed by the time of our observations, 
there are still probably numerous spots of emission in this narrow range of 
velocities that contribute to produce the two peaks in the spectra. This would 
explain why the fitted Gaussians peak velocity shifts are not real changes in 
velocity, but brought about by combination of masers that vary in intensity. 
The same limitations affect the four velocity peaks in the range $[-59,\ 
-55]$\,\kms; they are made up of $\sim30$ maser spots \citep{SVT11, HMW16} which 
have been fitted with only 7 or 8 Gaussians.

A noticeable feature of the polarization measurements is that they change 
from one pointing to the next. Based on our spider scans carried out as part 
of the process of calibrating the Mueller matrix \citep{FSG23}, this does 
not appear to be ``beam squint'', and therefore represents a time-varying 
measure of the polarization, rather than instrumental artifacts. A similar 
such variation is seen in the data for G213.705--12.60 which is briefly 
discussed later. 

During the four months during which these observations were made, the mean 
percentage of linear polarization across these spectra varied from 10\% to 
15 to 4 to 7\%. The peak linear polarization is at $v = -60.3$\,\kms, and 
changes from 11\% to 14 to 6 to 9\%, while the section between $v = -59$ and 
$-55.5$\,\kms\ initially has a negative slope, changing to a positive slope 
in Figure \ref{fig:G111_532}(b) and then back to a negative slope in Figure 
\ref{fig:G111_532}(f). The percentage circular polarization also undergoes 
changes from --10\% to +6\%, as can be seen in Figures \ref{fig:G111_526}(d) 
and (h) and \ref{fig:G111_532}(d) and (h). Zeeman splitting has been measured 
by fitting Gaussians to the RCP and LCP profiles, with LOS magnetic fields 
determined for the $-61.3$ and $-60.6$\,\kms\ profiles. These results are 
presented in Section \ref{sec:Z_split}. 

Overall, this is a complicated source that will need some detailed 
investigation. Our peak flux density is $S_{\nu} = 110$\,Jy with six clear 
peaks but lots of additional flux in the line wings. 

\begin{figure*}[ht!] 
\centering
\scalebox{0.6}{\plotone{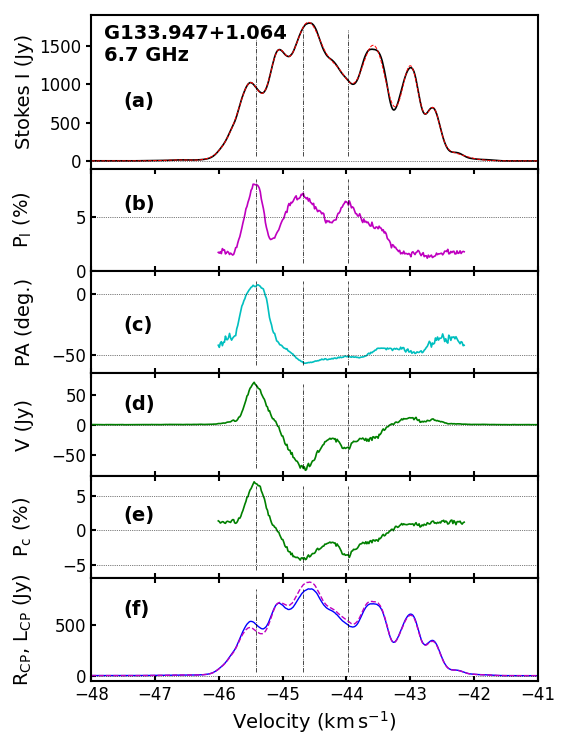}}
\caption{Spectra of G133.947+1.064 showing (a) Stokes $I$ (b) $P_\textrm{l}$, 
(c) PA of the linear polarization, (d) Stokes $V$, (e) $P_\textrm{c}$, and (f) 
the RCP and LCP components. The vertical dotted lines indicate peaks and/or 
dips in the $P_\textrm{l}$ and $P_\textrm{c}$ spectra.}
\label{fig:G133_7S}
\end{figure*}

\subsection{G133.947+1.064 (W3(OH))}
\label{sec:w3oh}
W3(OH) is a source rich in masers that has been observed extensively with 
both single-dish telescopes and interferometers at a range of frequencies. 
The 6.7\,GHz methanol masers were discovered by \citet{M91} within a velocity 
range of $[-48, -41]$\,\kms\ with a peak flux density $S_{\nu} = 3880$\,Jy. 
Linear polarization was found in seven maser spots with $S_{\nu} > 70$\,Jy by 
\citet{VHC06}. The mean percentage linear polarization was $\langle P_\textrm{l} 
\rangle = 1.8(7)$\% with a maximum value of 8\%. The error weighted position 
angle for these spots was $\langle \textrm{PA} \rangle = -67(9)$\degr. Zeeman 
splitting was detected by \citet{V08} using the 100m Effelsberg telescope. The 
peak flux density was 3705\,Jy and a Zeeman splitting of $\Delta V_\textrm{Z} 
= 0.141(3)$\,\ms\ at a velocity $v = -44.0$\,\kms\ was measured. \citet{SVL12} 
observed in full polarization mode using the EVN in 2009 November. They 
identified 51 maser spots in six regions. The peak flux density $S_{\nu} = 
2051$\,Jy had a velocity $v = -45.41$\,\kms. They found linear polarization in 
the range $P_\textrm{l} = [1.2,\ 8.1]\%$ and seven spots that displayed 
Zeeman splitting which varied by $\Delta V_\textrm{Z} = [-3.5,\ 3.8]$\,\ms. 
\citet{HMW16} found 38 spots of emission. The spectrum is complex and had a 
peak flux density $S_{\nu} = 2886$\,Jy. \citet{ASS23} found the masers 
displayed low variability over their five-year monitoring period. 

Our Stokes $I$ spectrum is shown in Figure \ref{fig:G133_7S} and is modeled 
using 10 Gaussians. However, due to the numerous overlapping maser profiles, 
these fits cannot be related to any particular spots. As in previous studies, 
the dominant maser emission occurs over the range $v= [-46, -42]$\,\kms and 
both linear and circular polarization can be measured in this region. As can 
be seen in Figure \ref{fig:G133_7S} the linear polarization varies between
1 and 8\%. The strongest polarization peak has a PA of 8\degr, while the
rest of the spectrum lies around a PA of $\sim-50\degr$. There is circular 
polarization across much of the profile, peaking at 7\% and then dropping to 
$\sim-4\%$. The dotted lines in the Figure indicate where the linear 
polarization has peaks -- these coincide with the circular polarization peak 
at $v= -45.4$\,\kms, and dips at the other two positions. These peaks and dips 
do not coincide with peaks in the flux density, which shows that interferometric 
observations are necessary to interpret which spots are polarized. No 
credible Zeeman splitting values were estimated from the fits to the RCP and 
LCP, and, therefore, we cannot compare our results with the results found 
by \citet{V08} or \citet{SVL12}.

\subsection{IRAS 05137+3919 (G168.063+0.820)}
\citet{XLH08} were the first to report 6.7\,GHz masers in this source. They 
found a peak flux density of 1.5\,Jy at $v = -16.2$\,\kms\ and another weak 
peak of 0.35\,Jy at $v = -20.7$\,\kms\ when they used the Effelsberg 100m 
telescope to observe in 2006 May. Earlier observations on the same telescope 
were made in May 2003 by \citet{FCS10} who also found a maser at $v = 
-16.2$\,\kms\ but with a peak flux density of 0.3\,Jy. Clearly, this maser 
had grown slightly in the intervening years. iMet have monitored this source 
for several years. They show a light curve for the $v = -18.66$\,\kms\ channel, 
changing to $v = -18.73$\,\kms\ midway through 2020, which does not correspond 
to either of the velocities found in earlier studies.

\begin{figure*}[ht!]
\centering
\scalebox{0.6}{\plotone{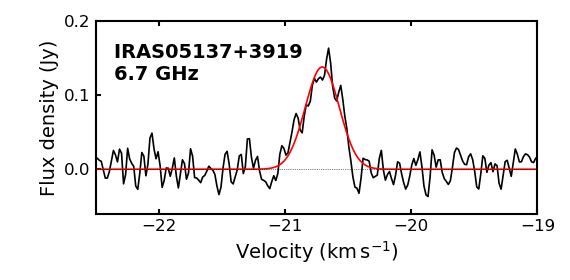}}
\caption{Stokes $I$ spectrum of IRAS 05137+3919.}
\label{fig:I05137}
\end{figure*}

We detected a single maser line at $v = -20.7$\,\kms\ with $S_{\nu} = 0.14$\,Jy, 
which has the same velocity as one of the lines found by \citet{XLH08}. Our 
Stokes $I$ spectrum of this feature is shown in Figure~\ref{fig:I05137} and is 
well-fitted by a single Gaussian profile. The line at $v = -16.2$\,\kms\ is no 
longer detectable to a limit of $\sim 26$\,mJy. Because of the small SNR, it is 
not surprising that no sign of any polarization was found.  

\begin{figure*}[ht!]
\centering
\scalebox{0.6}{\plotone{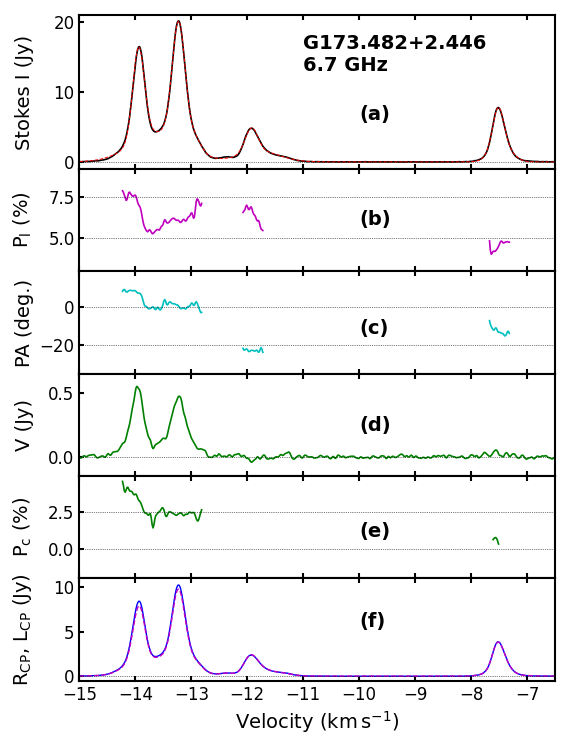}}
\caption{Spectra of G173.482+2.446 showing (a) Stokes $I$, (b) $P_\textrm{l}$, 
(c) PA of the linear polarization, (d) Stokes $V$, (e) $P_\textrm{c}$, and (f) 
the RCP and LCP components.}
\label{fig:G173_7}
\end{figure*}

\subsection{G173.482+2.446 (S231)}
This source was one of the 6.7\,GHz methanol masers discovered by \citet{M91}. 
It had a maximum flux density of 208\,Jy with masers predominantly lying between 
--14 and --13\,\kms. Zeeman splitting with $\Delta V_\textrm{Z} = 0.95(11)$\,\kms\ 
at $v = -13.0$\,\kms\ was found by \citet{V08} using the 100m Effelsberg telescope. 
\citet{SVL13} observed in full polarization spectral mode using the EVN in 2011 
October. They identified 32 maser spots, the strongest of which had a flux density 
$S_{\nu} = 23.419$\,Jy, at $v = -12.96$\,\kms\ with $P_\textrm{l} = 4.0(4)\%$ and 
$\textrm{PA} = 48(1)\degr$. They measured linear polarization of $0.8\% < P_\textrm{l} 
< 11.3\%$ in five spots but no circular polarization. 

By 2012 March when \citet{HMW16} used the VLA in the C configuration to look at 
this source, 24 spots of methanol masers were identified but the fluxes and 
velocities of the strongest sources had changed from those of \citet{SVL13}. 
According to \citet{ASS23} the masers display moderate variability. 

Our spectra are shown in Figure~\ref{fig:G173_7}. The four prominent peaks have 
narrow FWHMs (see Table \ref{tab:GaussFit}), suggesting single maser spots with 
polarization differing between the four features. Note that the Stokes $V$ 
signal for the peaks at $v = -13.94$ and --13.23\,\kms\ arise from a difference in 
the amplitude of the RCP and LCP spectra, whereas for the other profiles the 
amplitudes are the same, thereby producing $V \sim 0$. Gaussian fitting of RCP and 
LCP spectra in the $v = -13.94$ and --7.52\,\kms\ features indicate Zeeman 
splitting as presented in section \ref{sec:Z_split}. 

\subsection{IRAS 05382+3547 (G173.69+2.87)}
The 6.7\,GHz masers were discovered by \citet{SHK00} with a peak flux density of 
$S_{\nu} = 7.5$\,Jy at $v = -24.1$\,\kms. In the catalogue of \citet{SWB12} there 
is a single line at $v =-23.8$\,\kms\ with a flux density $S_{\nu} = 0.9$\,Jy. 
The source was monitored by iMet up to the end of 2023, showing emission with a 
flux density of $\sim1$\,Jy at $v = -24.1$\,\kms.

\begin{figure*}[ht!]
\centering
\scalebox{0.6}{\plotone{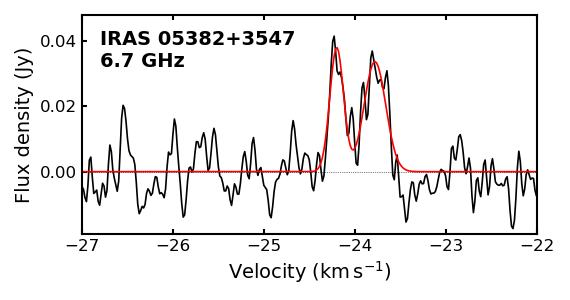}}
\caption{Stokes $I$ spectrum of IRAS 05382+3547.}
\label{fig:I05382}
\end{figure*}

\begin{figure*}[ht!]
\centering
\scalebox{0.6}{\plotone{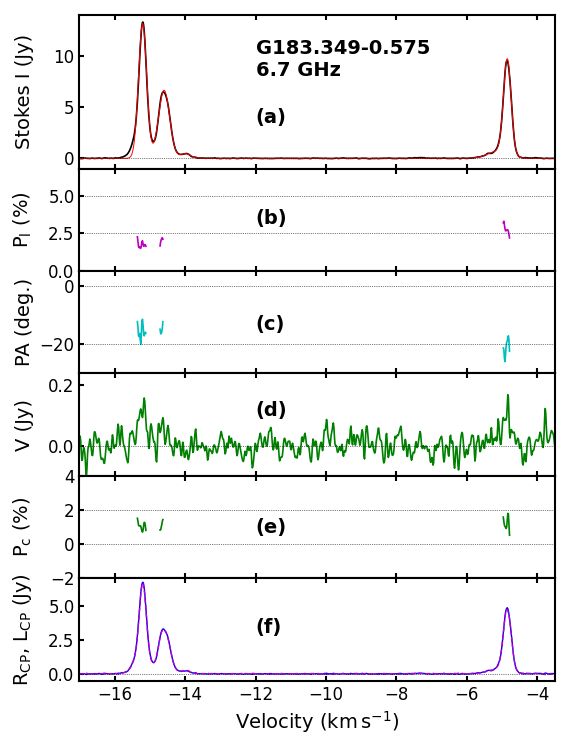}}
\caption{Spectra in G183.349--0.575 showing (a) Stokes $I$, (b) $P_\textrm{l}$, 
(c) PA of the linear polarization, (d) Stokes $V$, (e) $P_\textrm{c}$, and (f) 
the RCP and LCP components.}
\label{fig:G183_7}
\end{figure*}

We detected two masers with flux densities of 38 and 34\,mJy at $v = -24.2$ and 
$-23.8$\,\kms\ as shown in Figure \ref{fig:I05382}. This detection is at a $3\sigma$ 
level. There is a suggestion of Stokes $U$ and $V$ polarization, but with detection 
levels of $2\sigma$ this is not a definitive detection, so we have not presented it. 

\subsection{G183.349--0.575 (IRAS 05480+2545)}
Discovered by \citet{SVK99} using the Medicina 32 m telescope, the spectrum had two 
peaks each with a flux density $S_{\nu} = 3$\,Jy at $v = -15$ and $-5$\,\kms. In 
the catalogue of \citet{SWB12} a peak with $S_{\nu} = 13$\,Jy occurs at $v = 
-4.89$\,\kms\ and two peaks of 5 and 3.9\,Jy are around $v = -15$\,\kms. The VLA 
observations of \citet{HMW16} identified 19 spots of maser emission, 8 of which 
had velocities in the range $[-5.62, -4.39]$\,\kms, and 11 spots scattered between 
$[-15.28, -13.70]$\,\kms. The spots were clustered together in two different regions. 
The iMet monitoring data show that the peak at $v =-4.85$\,\kms\ is the strongest 
feature until about mid-2020 when the $v = -15.18$\,\kms\ feature starts to dominate. 

\begin{figure*}[ht!]
\centering
\scalebox{0.7}{\plotone{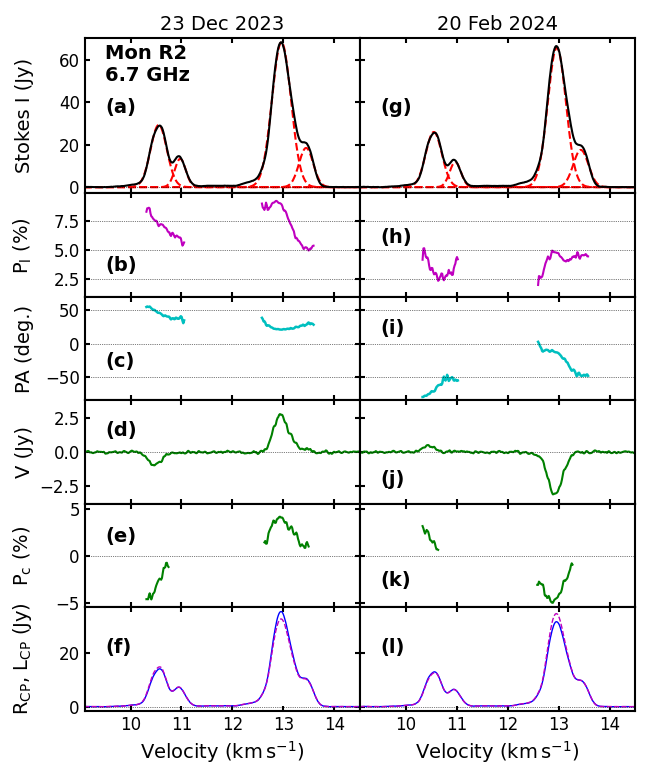}}
\caption{Spectra of G213.705--12.60 at the two epochs listed in Table \ref{tab:source} 
showing (a) Stokes $I$ (including dashed red lines for the four prominent maser 
profiles), (b) $P_\textrm{l}$, (c) PA of the linear polarization, (d) Stokes $V$, 
(e) $P_\textrm{c}$, and (f) the RCP and LCP components.}
\label{fig:Mon R2}
\end{figure*}

Our spectrum, shown in Figure~\ref{fig:G183_7}, has the strongest peak of 
13.16\,Jy at $v = -15.21$\,\kms, the second of 9.25\,Jy at --4.85\,\kms\ 
and the third peak has 6.66\,Jy at --14.60\,\kms. There are two weaker 
features at $v = -13.99$ and --5.11\,\kms. Our limited number of Gaussian 
fits to the profiles do not model the large number of spots found by 
\citet{HMW16}. What is new, however, are the polarization data presented 
here, which are the first for this source. There is both linear and circular 
polarization but only above $3\sigma$ levels in small regions around the 
major peaks. Within error ranges, there is no variation in the polarization 
properties found in the three main maser lines. Gaussian fitting of RCP and 
LCP suggests Zeeman splitting, but with low detection levels so that further 
observations are required for a definitive detection.   

\subsection{G213.705--12.60 (Mon R2)}
Methanol masers with a range of velocities between $v = 9 - 14$\,\kms\ and a 
peak flux of $S_{\nu}= 160$\,Jy were discovered by \citet{M91}. The spectrum 
changes in amplitude but the general shape consisting of four prominent peaks 
has remained similar over time. Monitoring at iMet and the Hartebeesthoek 
Radio Astronomy Observatory (G. MacLeod, private communication) show drifts 
in the velocities of the main peaks at $v = 10.6$ and 13.0\,\kms. EVN 
interferometric observations by \citet{SVL15} detected 20 maser features 
in an approximate northeast to southwest $\sim450$\,au linear distribution. 
They observed six masers with linear polarization fractions $P_\textrm{l}$ in 
the range of 3.0 to 5.0\% and PAs for five spots from 15(6)\degr\ to 23(3)\degr\ 
and PA = 51(49)\degr\ for the spot with $v = 10.68$\,\kms. Circular polarization 
with Zeeman splitting $\Delta V_\textrm{Z} = -6.6 \pm 1.0$\,\ms\ was found only 
in the brightest spot which had a velocity $v = 12.57$\,\kms. 

We have 18 observations of this source made between 2021 January and 2025 May, 
only two of which are presented here as they show a significant change in 
$P_\textrm{l}$ and PA between the two observations (Figure \ref{fig:Mon R2}). 
Further polarization changes are observed in the other observations with 
similar changes noted in the G213.705--12.60 4.765\,GHz exOH maser \citep{FS24}. 
This time series will be reported separately. In addition, the Stokes $V$ 
components have opposite direction for the two prominent masers and are seen 
to switch direction between the two epochs. Zeeman splitting is calculated for 
the prominent maser lines, and is seen to change direction in line with the 
change in circular polarization (see sections \ref{sec:Z_split} and \ref{sec:mag_fields}). 

\subsection{G189.030+0.783}
\citet{CVE95} discovered the masers in this source using the Parkes 64m
telescope. The spectrum contained two lines with flux densities $S_{\nu} 
= 17$ and 10\,Jy at velocities $v = 9$ and 10\,\kms. In the catalog of
\citet{SWB12} there are three lines, one of which has a clear shoulder 
indicating at least one more spot of emission. This region has been 
observed by \citet{GCF12} as part of the Methanol MultiBeam survey which 
reveals two nearby sources of maser emission, one of which is G189.030+0.783 
with two lines in the spectrum peaking at 16\,Jy, and the other source is 
\object{G188.946+0.886} which has its main line peaking at 607\,Jy at $v = 
10.8$\,\kms with a clear shoulder on the main peak. \citet{HMW16} found eight 
spots of emission, with a peak of 17.46\,Jy at $v = 8.96$\,\kms; all their 
other spots had flux densities $<10$\,Jy. In the iMet monitoring data for 
G189.030+0.783, the strongest flux density has a velocity $v = 10.83$\,\kms\ 
which is consistently larger than the flux density at $v = 8.93$\,\kms\ and
must be associated with G188.946+0.886 (see our G188.946+0.886 spectrum in Figure 
\ref{fig:G188_7} -- this feature is not in our G189.030+0.783 spectrum because 
the GBT has a smaller beam than iMet). G189.030+0.783 was classified as highly 
variable by \citet{ASS23}, but if the $v = 10.83$\,\kms\ line is excluded this 
source is not quite so variable.

\begin{figure*}[ht!]
\centering
\scalebox{0.6}{\plotone{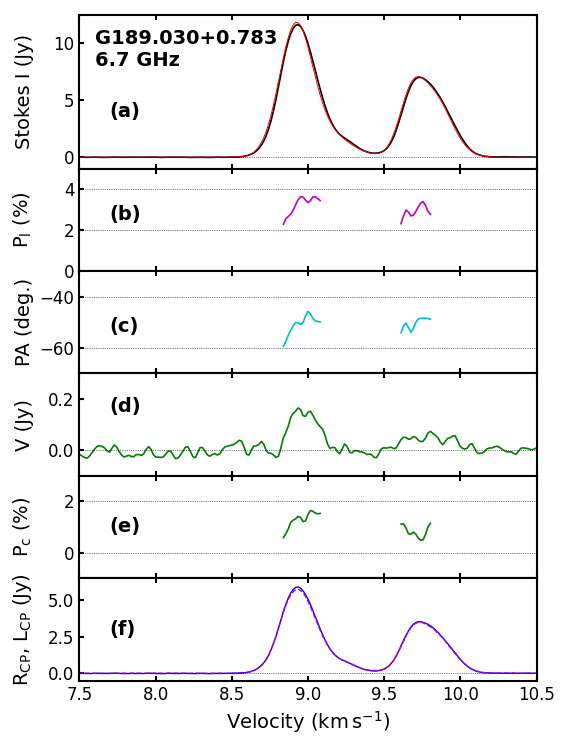}}
\caption{Spectrum of G189.030+0.783 showing (a) Stokes $I$, (b) $P_\textrm{l}$, 
(c) PA of the linear polarization, (d) Stokes $V$, (e) $P_\textrm{c}$, and 
(f) the RCP and LCP components}.
\label{fig:G189_7}
\end{figure*}

Our spectra of this source, which are the first to measure polarization, 
are shown in Figure~\ref{fig:G189_7}. Our $I$ spectrum is similar to that
of \citet{GCF12} and \cite{HMW16} in that we only have two broad peaks that
must be made up of multiple spots of emission and do not include the bright
maser from G188.946+0.886 in our beam side lobes. The brighter peak has a flux of 
10.7\,Jy, and we detected linear and circular polarization of a few percent 
across the spectrum. 

\subsection{G188.946+0.886 (IRAS 06058+2138)}
The 6.7\,GHz methanol masers in this source were found by \citet{M91} 
with a peak flux density $S_{\nu} = 457$\,Jy at $v = 8$\,\kms. A peak 
flux density of 633\,Jy was found by \citet{V08} using the 100m 
Effelsberg telescope, and a Zeeman splitting $\Delta V_\textrm{Z} = 
-0.49(2)$\,\kms\ at a velocity of 10.9\,\kms. The uncertainties were 
corrected by \citet{VTD11} to give a Zeeman splitting of $\Delta 
V_\textrm{Z} = -0.49(15)$\,\kms. The circular polarization measurements 
of \citet{V08} yielded a constant value for $B_{\parallel}$ across 
the spectrum from $v = 10.0 - 11.5$\,\kms. This source was observed again 
by \citet{SVL13} using the EVN in 2011 May 30, finding spots clustered in 
two groups. There were 36 spots in group A and 3 in group B. Linear 
polarization was measured in 11 spots with $P_\textrm{l}$ varying between 
1.3 -- 9.2\%, and circular polarization in one of these spots. 
Zeeman splitting $\Delta V_\textrm{Z} = 3.8(6)$\,\ms\ was measured in 
the brightest spot only with $P_\textrm{c} = 0.3$\%. \citet{HMW16} 
reported 25 spots of maser emission in a velocity range 
$v = [7.55, 12.29]$\,\kms. The two dominant velocities 
monitored by iMet occur at $v = 10.89$ and  10.51\,\kms. 
The overall shape of the spectrum has not changed significantly over the 
time period monitored by iMet, but the amplitude of the peak does vary.
These masers were found by \citet{ASS23} to be highly variable. 

\begin{figure*}[ht!]
\centering
\scalebox{0.6}{\plotone{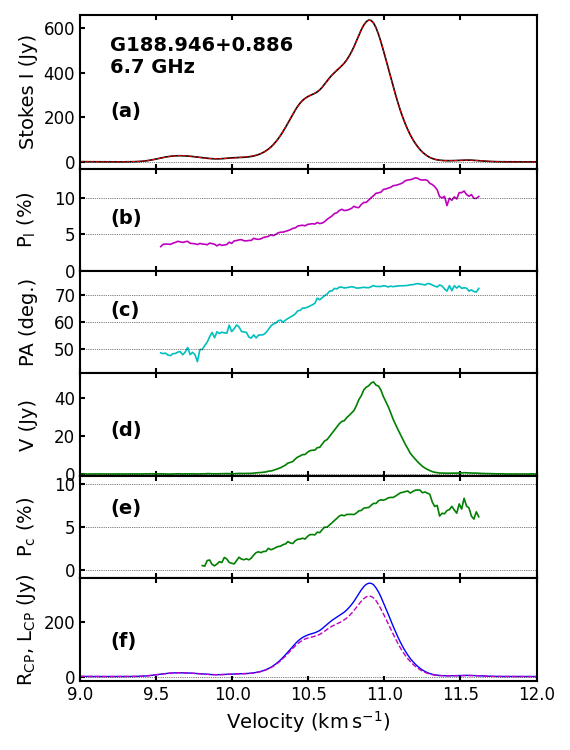}}
\caption{Spectra of G188.946+0.886 showing (a) Stokes $I$, (b) $P_\textrm{l}$, 
(c) PA of the linear polarization, (d) Stokes $V$ (e) $P_\textrm{c}$, and (f) the
RCP and LCP components.}
\label{fig:G188_7}
\end{figure*}

The spectra shown in Figure \ref{fig:G188_7} display linear and circular 
polarization which vary across the emission zone. The $I$ spectrum was 
fitted with 10 Gaussians which gives a good fit. The linear polarization 
increases from 3.4\% to 12.6\% with a PA that changes from 50\degr\ to 
75\degr. The \citet{SVL13} values of $P_\textrm{l}$ are not too different 
from what we get and our PAs are in a similar range to theirs. \citet{SVL13} 
found weak ($P_\textrm{v} = 0.3$\%) circular polarization in only one spot. 
The profile of our $P_\textrm{c}$ spectrum occurs over the same velocity 
range as magnetic fields reported by \citet{V08}, but increases from 0 to 
9.4\% as can be seen in Figure \ref{fig:G188_7}. Our circular polarization 
observations differ significantly from the values reported by 
\citet{SVL13}, but are not comparable because their reporting convention 
\citep[e.g.,][]{VDvL01,SVL22} differs from our definition for 
$P_\textrm{c}$ (refer to Appendix \ref{App:CirPol} for further details).

Gaussian fitting of RCP and LCP components indicates Zeeman splitting in two 
profiles. There is a separation of $\Delta V_\textrm{Z} = 3.0(4)$\,\ms\ at 
$v = 10.6867(6)$\,\kms, and $\Delta V_\textrm{Z} = 1.6(6)$\,\ms\ at $v = 
10.8800(7)$\,\kms\ (Table~\ref{tab:Mag_field_others}). This latter feature 
is close to that found by \citet{SVL13} ($v = 10.86$\,\kms), has a similar 
FWHM ($v = 0.24$ versus 0.28\,\kms), and is the brightest feature in the 
spectrum. The Zeeman splitting has changed from $\Delta V_\textrm{Z} = 
-0.49(15)$\,\ms\ in 2007 Nov \citep{VTD11}, to 3.8(6)\,\ms\ in 2011 May 
\citep{SVL13}, to 1.6(6)\,\ms\ in 2023 when our GBT observations were made. 
A feature at $v = -10.69$\,\kms, or Zeeman splitting in this velocity range, 
have not been reported previously. Magnetic fields associated with these 
splittings are discussed in section \ref{sec:mag_fields}.

\subsection{G196.454--1.677}
Masers were found by \citet{M91} with a peak flux density of 134\,Jy. Subsequently, 
numerous studies have been made of G196.454--1.677, which is also called \object{S269}, 
all of which show that the flux density varies dramatically on time scales of months 
to years. \citet{ASS23} classified this source as highly variable over their five 
yr monitoring program. In the \citet{SWB12} catalog, the peak of 15.6\,Jy occurs 
at $v = 15.14$\,\kms. \citet{HMW16} identified 14 spots over a velocity range of 
$v = [13.70, 16.16]$\,\kms\ with the peak of $S_{\nu} = 18.73$\,Jy at $v = 15.28$\,
\kms. Monitoring at iMet since 2013 shows the peak at $v= 14.70$\,\kms, shifting to 
14.71\,\kms after 2020, varying between $\sim5$ and $\sim30$\,Jy. The velocity 
spread of the emission appears to have remained steady over the velocity range 
$v = [13.7, 16.2]$\,\kms. 

\begin{figure*}[ht!]
\centering
\scalebox{0.57}{\plotone{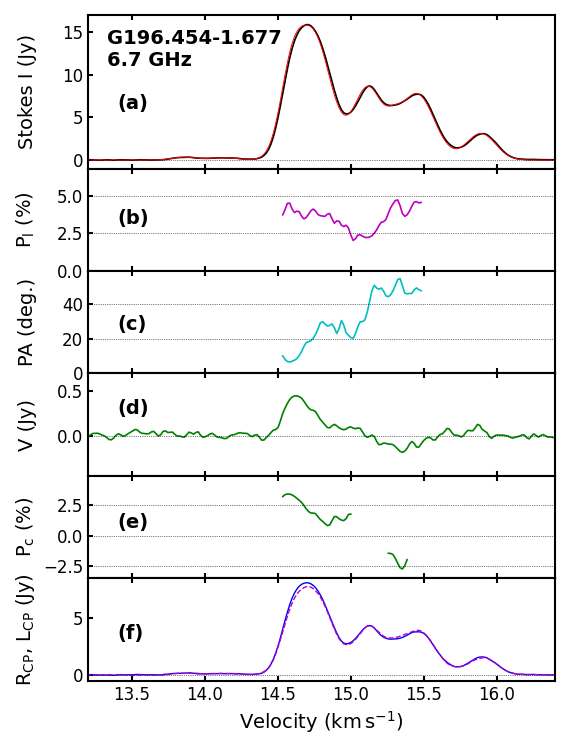}}
\caption{Spectra of G196.454--1.677 showing (a) Stokes $I$, (b) $P_\textrm{l}$, 
(c) PA of the linear polarization, (d) Stokes $V$ (e) $P_\textrm{c}$, and (f) the
RCP and LCP components.}
\label{fig:G196_7}
\end{figure*}

Our $I$ spectrum, shown in Figure~\ref{fig:G196_7}, was fitted with eight 
Gaussians leaving residuals of {$\pm 0.2$}\,Jy. Even though this is a good fit, 
narrow FWHM of two of the profiles (Table \ref{tab:GaussFit}) and \citet{HMW16} 
identifying 14 masers indicate interferometic observations are required to 
understand this source properly. Because G196.454--1.677 is a relatively weak 
source, it is not surprising that there is no previously reported polarization. 
We found linear polarization between 2.5 to 5\% and circular polarization at 
levels between --2.5\% and +3.0\%, lying between $v = [14.5, 16.0]$\,\kms, as can 
be seen in Figure~\ref{fig:G196_7}. The PA varied between 15\degr\ and 50\degr. 
Zeeman splitting has been determined for one component as discussed in 
section \ref{sec:Z_split}.

\subsection{G232.620+0.996 (IRAS 07299--1651)}
Maser emission was first detected by \citet{MG92} with a peak flux density $S_{\nu} 
= 42$\,Jy at a velocity $v = +23$\,\kms\ when observed in 1991 October. Since then 
there have been several single-dish observations showing variations in the flux 
density, but maintaining the overall spectral profile. VLA observations reported by 
\citet{HMW16} identified 15 spots of emission with velocities in the range $[21.42, 
24.06]$\,\kms. The strongest spot, with a flux density $S_{\nu} = 184.57$\,Jy, has 
a velocity $v = 22.83$\,\kms\ but a number of other spots are close enough in 
velocity to contribute to the unresolved peak seen in our single-dish spectrum. In 
the iMet monitoring data, the main peak has a velocity of 22.89\,\kms\ and reached a 
maximum of $\sim 350$\,Jy. The second peak has a velocity of 22.32\,\kms and flux 
density just below 100\,Jy that varies slowly. 

\begin{figure}[ht]
\centering
\scalebox{0.6}{\plotone{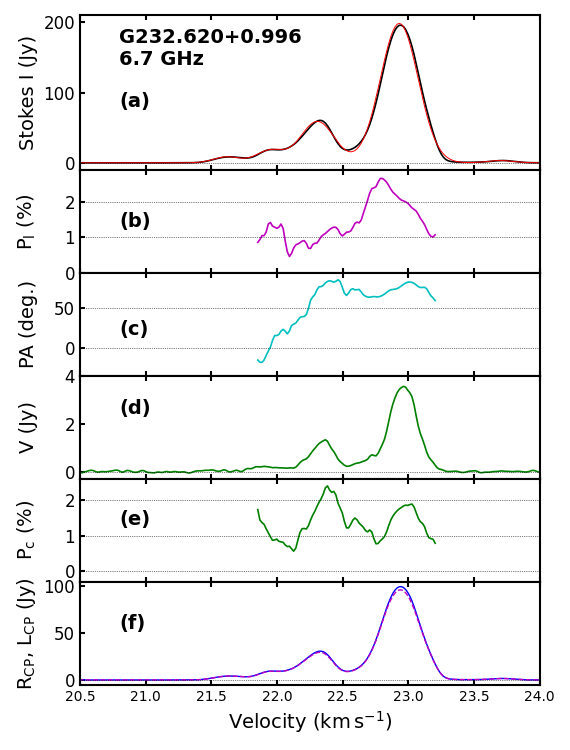}}
\caption{Spectra of G232.620+0.996 showing (a) Stokes $I$, (b) $P_\textrm{l}$, 
(c) PA of the linear polarization, (d) Stokes $V$ (e) $P_\textrm{c}$, and (f) the
RCP and LCP components.}
\label{fig:G232}
\end{figure}

Our polarization measurements are the first that we can find in the literature. 
The spectra of our observations are shown in Figure \ref{fig:G232}. This is the 
only one of our sources that has methanol but no exOH masers. There are several 
peaks in the single-dish spectrum, the strongest of which has a flux density of 
$S_{\nu} = 199$\,Jy at a velocity of $v = 22.93$\,\kms, but this is probably a 
combination of contributions from more than one spot. A five Gaussian fit to the 
$I$ spectrum left residuals of up to 10\,Jy, confirming that there are still a 
multitude of maser spots within the narrow range of velocities from this source 
that cannot be resolved by single-dish observations. 

Both linear and circular polarization were found in G232.620+0.996, which vary 
across the spectrum, again indicating different polarizations from masers in the 
same region. The dominant peak, at $v = 22.93$\,\kms, has linear polarization 
of $\sim 2.5\%$ at a PA of $\phi \sim 80\degr$, and circular polarization of 
$P_\textrm{c}\sim 1.8\%$. The peaks in the polarization spectra do not align 
with the peaks in the flux density, indicating the complexity due to multiple 
overlapping maser spots that contribute to the spectrum. Gaussian fitting of 
RCP and LCP indicates Zeeman splitting in one component as discussed further 
in section \ref{sec:Z_split}. 

\subsection{G240.316+0.071 (IRAS 07427--2400)}
This object has been observed on numerous occasions looking for 6.7\,GHz 
masers but all searches thus far have produced a null result. We obtained 
two observations of G240.316+0.071, detecting a feature at 
$v = 63.2$\,\kms\ on both occasions and a tentative feature at 
$v = 62.84$\,\kms\ during the first observation. The Gaussian 
parameters of the fitted profiles are listed in the table in the 
Appendix and spectra are shown in Figure \ref{fig:G240_7}. 

\begin{figure*}[ht!]
\centering
\scalebox{0.85}{\plotone{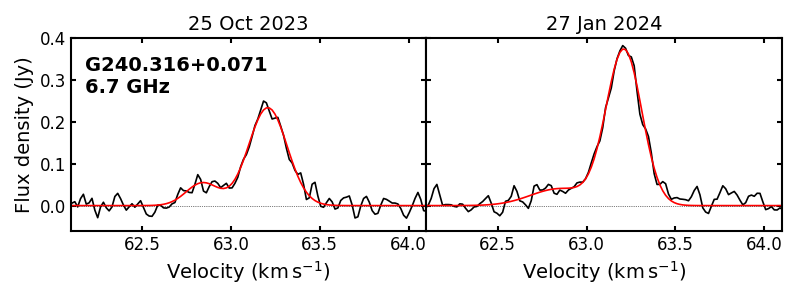}}
\caption{Stokes $I$ spectra of G240.316+0.071 on the two epochs listed in Table 
\ref{tab:source}.} 
\label{fig:G240_7}
\end{figure*}

The spectra could be fitted using two Gaussians with similar velocities on 
both dates. The dominant peak's flux density increased noticeably, from 234 to 
371\,mJy, during the three month interval between our observations. However, 
the maximum flux density of the tentative component at $v = 62.84$\,\kms\ 
declined. The masers are weak and show no detectable linear or circular 
polarization. 

Null detections of 6.7\,GHz in previous searches and the changes over a 
three month interval indicate that this is a rapidly varying source. It 
would be worthwhile to monitor this source regularly to see if the masers 
continue to grow or if they are just a transient feature. 

\section{Discussion}
Methanol maser spectra are generally a complex combination of maser 
spots with overlapping profiles that cannot be resolved with single-dish 
telescopes. Single maser profiles are observed but usually only when 
the source is weak. IRAS 05137+3919, IRAS 05382+3547 and G240.316+0.071 can 
all be fitted with one or two Gaussians but have flux densities $<1$\,Jy. 
The Gaussian fitting described in this paper can decompose spectral 
profiles, although without interferometric observations it is not possible 
to confirm the Gaussian profile as a single maser spot. Due to flux variability 
in most of these sources, new spots arise and old ones quench, 
so that archival interferometric observations are of limited use in 
identifying velocity components in our spectra.

Most of the methanol spectra display both linear and circular polarization. 
Linear polarization is observed in the range of 0 to 15\%, and circular 
polarization in the range of 0 to $\pm9\%$. Our sources with no
polarization detection have low flux densities, and polarization of a few 
percent would lie in our rms noise and, if it is present, is thus not 
detected. Based on our limited sample of sources, it is expected that a 
significant fraction of known 6.7 GHz maser sources will show polarized 
emission and, therefore, a larger observational campaign is required.

\subsection{Zeeman splitting and linear polarization comparisons} 
\label{sec:Z_split}
The process of fitting Gaussian profiles to both RCP and LCP spectra enables 
determining the velocity separation due to Zeeman splitting for distinct 
peaks in some of the spectra. For the sources in Tables \ref{tab:Mag_field_G111} 
and \ref{tab:Mag_field_others} the fitted Gaussian profile could represent more 
than one maser spot despite it appearing as a distinct peak in the spectrum. 
These Zeeman splitting values are thus a weighted average for maser spots that 
form the Gaussian profile. The four prominent profiles in G213.705--12.60 
(Table \ref{tab:Mag_field_MonR2}) represent distinct maser spots, as confirmed 
by e-MERLIN observations in 2024 October and March 2025 (Fallon \& Smits, 
in preparation). The polarization values, also presented in Tables 
\ref{tab:Mag_field_G111} and \ref{tab:Mag_field_MonR2}, are average values for 
a window 0.06\,\kms\ either side of the central velocity. The window covers 
$\sim8$ channels, and the error value is the standard deviation over the 
individual channel values.

Differences in Zeeman splitting values are seen for maser profiles in the same 
spectrum. The two profiles in NGC 7538 have different splitting values (Table 
\ref{tab:Mag_field_G111}), and significant Zeeman splitting differences are seen 
between the maser profiles in G213.705--12.60 (Table~\ref{tab:Mag_field_MonR2}). 
Both these sources show Zeeman splitting changing between the observation epochs, 
with the Zeeman splitting in G213.705--12.60 switching direction in line with 
the change in circular polarization. 

Comparisons of Zeeman splitting values from previous studies highlight similarities 
and differences. NGC7538 was observed by \cite{SVT11} who measured $\Delta 
V_\textrm{Z} = 2.7(3)$\,\ms\ at $v = -60.49$\,\kms. This is similar to some of our 
values for the $-60.6$\,\kms\ profile presented in Table \ref{tab:Mag_field_G111}. 
Their $P_l =  4.5(8)$ is slightly below the lower range of values that we measured. 
\cite{V08, VTD11} report $\Delta V_\textrm{Z} = 0.79(3)$\,\ms\ but for a different 
part of the spectrum, with slight differences seen in a second set of observations 
done nine months later. 

In G173.482+2.446 \cite{V08} found Zeeman splitting $\Delta V_\textrm{Z} = 
0.95(11)$\,\ms\ at $v = -13.0$\,\kms. We measure $\Delta V_\textrm{Z} 
= -0.9(1)$\,\ms\ at $v = -13.9$\,\kms\ (see Table \ref{tab:Mag_field_others}), 
which has a similar magnitude but opposite direction. 

For G188.946+0.886 \cite{VTD11} report $\Delta V_\textrm{Z} = -0.49(15)$\,\ms\ 
at $v = 10.9$\,\kms. At $v=10.88$\,\kms\ we measure $\Delta V_\textrm{Z} = 
1.6(6)$\,\ms, which has both a different magnitude and direction. \cite{SVL13} 
measured a Zeeman splitting $\Delta V_\textrm{Z} = 3.8(6)$\,\ms\ for the bright 
maser which was at $v = 10.86$\,\kms.  They report $\textrm{PA}=76(5)\degr$ for 
this feature which is similar to our value shown in Figure~\ref{fig:G188_7}. 
However, their $P_l = 6.0(1.3)\%$ differs from our values (Figure~\ref{fig:G188_7}). 

A value of $\Delta V_\textrm{Z} < 0.77$\,\ms\ at $v = 15.0$\,\kms\ for 
G196.454--1.677 is reported by \cite{VTD11}. This is different to the $\Delta 
V_\textrm{Z} = -1.8(7)$\,\ms\ that we measure at $v = 15.1$\,\kms\ (see Table 
\ref{tab:Mag_field_others}). 

For G213.705--12.60 \cite{SVL15} report $\Delta V_\textrm{Z} = -6.6(10)$\,\ms\ 
at $v = 12.57$\,\kms\ which is more than double the values we measure for the $v = 
12.96$\,\kms\ maser profile (Table\ \ref{tab:Mag_field_MonR2}). Note these are 
believed to be the same maser as this profile exhibits velocity drift (G. MacLeod, 
private communication). The $P_l = 4.6(0.4)\%$ and $\textrm{PA} = 16(7)\degr$ 
reported by \cite{SVL15} are similar to the range of values we measure.

These comparisons show the linear polarization and Zeeman splitting in some masers 
changing between observations. There are no published results of regular monitoring 
of maser polarization, and only limited commentary on changes in polarization or 
Zeeman splitting for some masers. Monitoring of methanol maser flux density using 
single-dish telescopes is done on a regular basis, but monitoring of the maser 
polarization has not been undertaken. Single-dish monitoring is more feasible than 
interferometric monitoring, and can be valuable in tracking maser polarization 
behavior and changes in Zeeman splitting. Furthermore, single-dish maser 
polarization monitoring can provide direction as to when interferometric 
observations might be justified.

\begin{table*}
   \centering
    \caption{Zeeman splitting, magnetic field and polarization values 
          for the two distinct maser profiles of NGC 7538 (G111.526+0.803 (526), 
          G111.532+0.759 (532) and G111.542+0.777 (542)).}
   \label{tab:Mag_field_G111}
   \begin{tabular}{cll*8{c}}
   \hline 
No. &Source  &Observation &Flux     &Velocity &FWHM   &Zeeman     &Magnetic   &$P_\textrm{c}$ &$P_\textrm{l}$ &PA \\
    &Name    &date in     &Density  &         &       &splitting  &field      &    &   &   \\
    &G111.\# &2023        &(Jy)     &(\kms)   &(\kms) &(\ms)      &(mG)       &(\%)     &(\%)  &(degrees)\\ 
\hline
2a &526  &Apr 21 &15.73(3)  &--61.2860(3) &0.3665(8)  &1.2(3)   &--24(5)   &0.3(2)  &8.1(2)   &17.1(8)\\
2b &     &Jun 23 & 8.26(2)  &--61.3043(4) &0.3656(10) &3.3(14)  &--65(27)  &--5(2)  &11.4(15) &17.3(14)\\
3  &532  &Jul 07 &38.66(6)  &--61.2903(3) &0.3664(8)  &--0.6(3) &12(5)     &3.4(2)  &3.4(2)   &85(2)\\
4  &542  &Aug 03 &83.04(13) &--61.3084(3) &0.3669(8)  &1.7(2)   &--34(4)   &1.0(2)  &7.16(8)  &--30.3(15)\\  
\hline
2a &526  &Apr 21 &13.10(2)  &--60.6383(5) &0.4776(12)  &3.1(3)   &--60(7)  &2.5(3)  &10.6(7) &25.5(12)\\
2b &     &Jun 23 & 7.00(2)  &--60.6560(6) &0.4737(15)  &3.3(19)  &--64(36) &--2(7)  &14(3)   &23(2)\\
3  &532  &Jul 07 &32.86(6)  &--60.6404(4) &0.4730(12)  &--0.4(3) &8(6)     &1.4(4)  &5.5(5)  &68(2)\\ 
4  &542  &Aug 03 &70.90(12) &--60.6578(4) &0.4726(11)  &3.4(3)   &--67(6)  &4.8(4)  &8.6(5)  &--15.9(19)\\  
\hline
\end{tabular}
\end{table*}

\begin{table*}
\centering
\caption{Gaussian parameters, Zeeman splitting and magnetic field values from 
         distinct profiles with apparent single masers.}
\label{tab:Mag_field_others}
\begin{tabular}{c*7{c}}
\hline
No. &Source &Gaussian      &Flux      &Velocity &FWHM   &Zeeman    &Magnetic \\
    &Name   &Number from   &Density   &         &       &splitting &field \\
    &       &Table 5       &(Jy)      &(\kms)   &(\kms) &(\ms)     &(mG)\\
\hline
12 &G173.482+2.446  &1    &13.29(2)  &--13.9420(2)    &0.2456(5)  &--0.9(1) &17(2) \\
   &                &4    &5.28(13)  &--7.5164(7)     &0.229(3)   &--1.3(5) &25(10) \\
\hline
15 &G188.946+0.886  &4    &234(3)   &10.6867(6)      &0.1987(13) &3.0(4)   &--58(8) \\
   &                &5    &394(39)  &10.8800(7)      &0.241(4)   &1.6(6)   &--32(12) \\
\hline
17 &G196.454--1.677  &5  &8.37(2)  &15.1184(5)      &0.2333(18) &--1.8(7) &36(13) \\
\hline
19 &G232.620+0.996  &4  &198.7(2)  &22.9306(2)      &0.3377(4)  &0.4(1)   &--7(2) \\
\hline
\end{tabular}
\end{table*}

\begin{table*}
\centering
\caption{Gaussian parameters, Zeeman splitting, magnetic field and polarization values 
for the four prominent maser peaks in G213.705--12.60. The fitting process uses seven Gaussian 
profiles for each epoch.}
\label{tab:Mag_field_MonR2}
\begin{tabular}{lcccccccc} 
\hline
Date       &Flux Density  &Velocity  &FWHM   &Zeeman split &Magnetic   &$P_\textrm{c}$ &$P_\textrm{l}$ &PA \\
observed   &(Jy)          &(\kms)    &(\kms) &(\ms)        &field (mG) &(\%)      &(\%)  &(degrees)\\
\hline
2023 Dec 23  &29.31(10)  &10.5536(5)  &0.3726(15)  &2.7(4)   &--53(7) &--2.7(3)   &7.2(2) &46(1) \\
2024 Feb 20  &25.98(10)  &10.5514(5)  &0.3686(16)  &--2.1(4) &41(8)   &1.4(3)     &3.2(3) &--71(2) \\
\hline
2023 Dec 23  &13.21(13)  &10.9728(9)  &0.248(3)    &         &        &           &5.9(3) &39(3)\\
2024 Feb 20  &11.68(13)  &10.9701(10) &0.252(3)    &         &        &           &3.8(7) &--54(4)\\
\hline
2023 Dec 23  &68.50(7)  &12.9606(3)  &0.4281(9)  &--0.9(2)  &18(4)    &4.1(1)     &8.8(3) &21.2(4) \\
2024 Feb 20  &66.64(7)  &12.9651(3)  &0.4281(9)  &2.5(2)    &--48(4)  &--4.5(2)   &4.8(1) &-13.1(9)\\
\hline
2023 Dec 23  &18.74(7)  &13.4374(10)  &0.342(2)  &--3.4(8)  &67(15)   &1.1(5)     &5.1(2) &29(3)\\
2024 Feb 20  &17.49(7)  &13.4402(11)  &0.344(2)  &3.4(8)    &--66(15) &--0.1(5)   &4.5(2) &-46(3)\\
\hline
\end{tabular}
\end{table*}

\subsection{Magnetic fields} \label{sec:mag_fields}
The 6.7\,GHz methanol transition comprises eight hyperfine components, each with 
different Land\'{e} $g$-factor contributions \citep{LVS18}. It is not clear which 
hyperfine transition, or combination of components, produce the Zeeman splitting. 
The level of maser excitation, and which hyperfine transition is dominant, will 
influence the polarization fraction and Zeeman splitting \citep{DVLS20}. From our 
calculated Zeeman splittings we estimate the LOS magnetic field using the Land\'{e} 
$g$-factor for the $F=3 \rightarrow 4$ hyperfine transition. This component has a 
Zeeman coefficient of $-1.135$\,Hz\,mG$^{-1}$ = $-0.05125$\,\ms\,mG$^{-1}$ 
\citep{LVS18}. It has the largest Land\'{e} $g$-factor value of the hyperfine 
transitions and thus our estimates of the magnetic field represent a lower limit. 
Magnetic field values estimated here are in the range of 10's of mG, derived 
from Zeeman splitting values in the range $\Delta V_\textrm{Z} = [-3.4, +3.4]$, 
which are similar to Zeeman splitting values reported by \citet{V08} and 
\citet{SVL22} for methanol maser observations. 

As our estimated LOS magnetic field values are proportional to the Zeeman
splitting, the same changes are seen, i.e.\ for NGC 7538 and G213.705--12.60 
the values differ between the masers profiles and are seen to change over time,
aligning with changes in the linear and circular polarization (see Tables 
\ref{tab:Mag_field_G111} and \ref{tab:Mag_field_MonR2}).  

\subsection{Possible reasons for changes in polarization, Zeeman splitting 
and magnetic fields}
\label{sec:pol_changes}
Variations in both linear and circular polarization are observed across the maser 
profiles of several sources, specifically NGC7538, G133.947+1.064, G173.482+2.446, 
G213.705-12.60, G188.946.886, G196.454-1.677, and G232.620+0.996 (see Figures 
\ref{fig:G111_526} -- \ref{fig:G133_7S}, \ref{fig:G173_7}, \ref{fig:Mon R2}, 
\ref{fig:G188_7} -- \ref{fig:G232}). Notably, G133.947+1.064, G213.705-12.60, and 
G196.454-1.677 exhibit profiles with opposing circular polarization signs (see 
Figures \ref{fig:G133_7S}, \ref{fig:Mon R2} and \ref{fig:G196_7}). Furthermore, 
multi-epoch observations of NGC7538 and G213.705-12.60 reveal temporal variability 
in both linear and circular polarization.

Several mechanisms have been put forward to explain differences in the 6.7\,GHz 
methanol maser polarization and Zeeman splitting \citep{DVS17, DVLS20, KSB25}. 
Different hyperfine components of the 6.7\,GHz transition could be dominant in 
different masers and change between epochs. Alternatively, the observed profiles 
are a blend of masers along the LOS with relative intensities that vary over time 
(presumably due to changing local conditions within the cloud). Another 
consideration is that the magnetic field direction and strength vary between 
masers, and as a function of time, producing the changing polarization 
characteristics that are observed.

The observed polarization changes in G213.705--12.60 between the two epochs 
(see Figure \ref{fig:Mon R2}) presents an opportunity to evaluate these 
plausible explanations. Based on the results presented by \cite{DVLS20}, 
an order of magnitude change in excitation temperature may be required for 
the observed change in linear polarization, while the circular polarization 
and Zeeman splitting switch in direction would require a different hyperfine 
transition to be dominant. However, the spectrum profile and flux densities 
do not change significantly between the two epochs. It seems unlikely that 
the conditions, such as density and temperatures, in the maser cloud would 
change sufficiently to result in this noticeably different polarization 
outcome. In addition, similar polarization changes are observed over the 
same period in the G213.705--12.60 4.765\,GHz exOH maser \citep{FS24, SF25}. 
Polarization in the 4.765\,GHz line is generated by a single hyperfine 
transition, $m_{F} = \pm1 \rightarrow 0$, so the changes cannot relate to 
different hyperfine components.

A more consistent view is that the polarization and Zeeman splitting 
variations arise from changes in the magnetic field threading the maser 
and varying over periods of months. It is quite possible that, for the other 
sources presented in this paper, changes in the magnetic field direction 
and/or strength produce the polarization variations seen in the maser profiles. 
Methanol maser spots observed with milliarcsecond resolution have shown  
different magnetic field orientations depending on their location \citep[e.g. ]
[]{SSM15}. Since maser spots with similar velocities are typically spatially 
clustered, this supports the view that distinct maser profiles within a 
single-dish spectrum can exhibit different polarization characteristics, and 
therefore different magnetic field orientations and strengths, as observed in 
this work.

Furthermore, the position of a maser can also change over time \citep{HRO23}, 
so it is not unreasonable that the observed magnetic field time variability is 
due to a proper motion shift in position of the maser, as opposed to a time 
varying magnetic field. Magnetic field behavior within the protostellar accretion 
disk is not well understood: there are theoretical studies and magnetohydrodynamic 
simulations, but only limited observations \citep{PR19}. Accretion disk theory 
assumes that turbulence in the disks will result in a viscus torque, and posits 
that the disk winds are magnetized. If these magnetized disk winds interact with 
the maser regions, it is possible that the magnetic field varies between regions 
of maser emission and changes with time. Further investigation is required to 
understand the spatial and time variations of magnetic fields in star-forming regions.

\section{Conclusions}
We present results from a small RA limited survey of 6.7\,GHz methanol masers 
made using the 100m GBT in full Stokes mode. Of our 21 pointings, the observations 
towards G111.526+0.803 (2a \& b), G111.532+0.759 (3) and G111.542+0.777 (4) can be 
counted as one source (NGC7538), leaving 19 sources. There were 6 non-detections 
of masers in these 19 sources, and three sources, IRAS 05137+3919, IRAS 05382+3547 
and G240.316+0.071, had weak masers with no polarization detectable above the noise. 
G240.316+0.071 is a new discovery of time-variable masers. Nine sources had linear 
and circular polarization. G108.758--0.986 was our only source with linear 
polarization, but no circular polarization above our detection limits.  

Circular polarization was detected in G173.482+2.446 for the first time, and the 
first polarization measurements were made on G108.758--0.986, G183.349--0.575, 
G189.030+0.783, G196.454--1.677 and G232.620+0.996. We had multiple observations 
towards NGC 7538 and G213.705--12.60 in which time-variable polarization was found 
in both sources. Most spectra showed a complex combination of overlapping maser 
profiles that cannot be resolved with single-dish telescopes, so interferometric 
observations are required to determine the properties of individual maser 
features, but single-dish observations can be used to identify flux density 
and polarization variations for follow-up observations with interferometers. 

Magnetic fields were estimated for some profiles in several of the sources 
using Gaussian fitting of the RCP and LCP spectra to determine Zeeman splitting. 
The magnetic fields have strengths of up to $\sim70$ mG which aligns with previous 
observations of magnetic fields estimated using methanol masers in star-forming 
regions. 

Zeeman splitting varies between the maser profiles and is seen to change 
over time in line with changes in the linear and circular polarization. 
A reasonable explanation for this is that the polarization and Zeeman 
splitting vary between masers in the same source due to changes in the 
magnetic field threading the masing gas and varying over periods of 
months, though maser proper motion may explain the time variability. 
These temporal changes are the reason why our results differ from previous 
measurements of the polarization characteristics in these sources, and 
indicate that regular monitoring of the polarization properties of 6.7\,GHz 
methanol masers is needed to get a better understanding of the changes.

\begin{contribution}
Observations and data analysis were undertaken by PF, whilst DPS completed 
the 6.7 GHz methanol maser literature survey and commentary on each of the 
sources. Discussion, conclusions and review of the paper are a combined 
effort.
\end{contribution}

\begin{acknowledgments}
This material is based upon work supported by the National Radio Astronomy 
Observatory and Green Bank Observatory which are major facilities funded 
by the U.S. National Science Foundation operated by Associated Universities, 
Inc.

The observations were part of GBT project AGBT22B-354. 
\end{acknowledgments}

\vspace{5mm}
\facilities{GBT}

\software{GBTIDL \citep{MGB13}, Matplotlib \citep{Hunter:2007}}

\bibliographystyle{aasjournalv7}
\bibliography{AASmeth}

\appendix
\section{Circular polarization definition}
\label{App:CirPol}
Equation (\ref{eq:CirPol}) is a standard definition of circular polarization 
$P_\textrm{c}$. We have noted that several authors use 
\begin{equation}
P_\textrm{V} = \frac{V_\mathrm{max} - V_\mathrm{min}}{I_\mathrm{max}} 
\end{equation}
as a definition of circular polarization when a characteristic S-shaped $V$ 
spectrum is found, where $V_\mathrm{max}$ and $V_\mathrm{min}$ are the 
maximum and minimum heights of the $V$ spectrum. $I_\mathrm{max}$ is the 
maximum intensity of the Stokes $I$ spectrum. 
The formula has been used for interpreting the circular polarization 
\citep{NW92}, and originates back to \citet{FG89} and \citet{GSS60} who use 
this to measure Zeeman separation. Therefore, our circular 
polarization values $P_\textrm{c}$ are not comparable to previously published 
results that use $P_\textrm{V}$.

In an attempt to reaffirm what $P_\textrm{V}$ measures, and to assign 
uncertainties to these values, we have used the two Gaussian model introduced 
by \citet{SF25} to analyze their circular polarization measurements of 
4.765\,GHz exOH masers to provide some insight. 

Let the RCP and LCP profiles be given by two Gaussians that are shifted in 
velocity by a Zeeman splitting of $\pm\delta$ from the central maser velocity 
(which is taken to be at $\mu = 0$), then 
\begin{eqnarray} 
\textrm{RCP} = A\,\me^{-(v -\delta)^2/2\sigma^2} \quad \mathrm{and} \quad 
\textrm{LCP} = A\,\me^{-(v+\delta)^2/2\sigma^2} \ , \label{eq:RCP}
\end{eqnarray}
and the Stokes $V$ component can be written as
\begin{eqnarray} 
V & = & A\left[ \me^{-(v -\delta)^2/2\sigma^2} - \me^{-(v+\delta)^2/2\sigma^2} 
\right] \\
   & = & A\,\me^{-v^2/2\sigma^2} \me^{-\delta^2/2\sigma^2} \left[ 
   \me^{v\delta/\sigma^2} - \me^{-v\delta/\sigma^2} \right]. \label{eq:V}
\end{eqnarray}
The derivative of $V$ with respect to velocity $v$ is then
\begin{eqnarray} 
\frac{\dif V}{\dif v} & = & A\,\me^{-v^2/2\sigma^2} \me^{-\delta^2/2\sigma^2} 
\left[ \frac{-v}{\sigma^2} \left(\me^{v\delta/\sigma^2} - \me^{-v\delta/\sigma^2} \right) 
+ \frac{\delta}{\sigma^2} \left(\me^{v\delta/\sigma^2} + \me^{-v\delta/\sigma^2} \right) 
\right]\ .
\end{eqnarray}

To find the velocities at which the maximum, $V_\textrm{max}$, and minimum, 
$V_\textrm{min}$, of $V$ occur,  set $\dif V/\dif v = 0$. This gives
\begin{equation} 
v\left(\me^{v\delta/\sigma^2} - \me^{-v\delta/\sigma^2} \right) = \delta \left(
\me^{v\delta/\sigma^2} + \me^{-v\delta/\sigma^2} \right) . 
\end{equation}

If $v\delta \ll \sigma^2$, a Taylor series to first order can be used to expand 
$\me^{-v\delta/\sigma^2} = 1 - v\delta/\sigma^2$ and $\me^{v\delta/\sigma^2} = 
1 + v\delta/\sigma^2$, and then substitute these into the above equation to get
\begin{equation}
v \left( \frac{2v\delta}{\sigma^2} \right) = 2\delta \quad \Longrightarrow \quad 
v = \pm \sigma \ . \label{eq:v=sigma}
\end{equation}

Applying this same approximation to equation (\ref{eq:V}), a Taylor expansion to 
first order gives 
\begin{equation}
 V = A \me^{-v^2/2\sigma^2}\left[ \left( 1 + \frac{v \delta}{\sigma^2}\right) 
 - \left( 1 - \frac{-v \delta}{\sigma^2} \right) \right] = A \me^{-v^2/2\sigma^2}
 \left[ \frac{2 \delta v}{\sigma^2} \right]  \ . 
\end{equation} 
The Stokes $I$ component is a sum of the RCP and LCP components and has a maximum 
value $I_\textrm{max} = 2A$ using this same approximation.

Using equation (\ref{eq:v=sigma}) for the velocities at $V_\textrm{max}$ and 
$V_\textrm{min}$, it is straight forward to show that 
\begin{equation} 
P_\textrm{V} = \frac{V_\textrm{max} - V_\textrm{min}}{I_\textrm{max}} = 
\frac{2 \me^{-1/2}\delta}{\sigma}  \ .
\label{eq:Vmax-Vmin/Imax} 
\end{equation}
Thus confirming $P_\textrm{V}$ as a measure of the Zeeman splitting 
($\delta$ is equivalent to $\Delta V_\textrm{Z}$ which is used by other 
authors). Uncertainties in the measured values of $V_\textrm{max}$, 
$V_\textrm{min}$, $I_\textrm{max}$ and $\sigma$ can be used to estimate 
an uncertainty for $\delta$.

\citet{SF25} showed that within the approximation $v\delta \ll \sigma^2$, $V/I$  
is a straight line with a slope $m = \delta/\sigma^2$ when the characteristic 
S-shaped $V$ is observed. To measure the Zeeman splitting, an alternative 
is to fit a straight-line to $V/I$ but, as pointed out by \cite{SF25}, fitting 
$\dif I/\dif v$ to $V$ results in even better measurement accuracy.

\section{Gaussian parameters}
\begin{longtable*}[c]{clcccc}
\caption{Gaussian parameters fitted to our Stokes $I$ spectra.}
\label{tab:GaussFit}
\tabularnewline\hline
No. &Source          &Gaussian     &Peak flux     &Velocity  &FWHM \\
    &                &number       &density (Jy)  &(\kms)    &(\kms)  \\ 
\hline
\endhead
1  &G108.758--0.986  &1  &0.09(2)  &--56.099(18)  &0.22(4)\\
   &                 &2  &0.16(1)  &--54.591(14)  &0.41(3)\\
   &                 &3  &0.07(1)  &--46.89(3)    &0.37(7)\\
   &                 &4  &0.53(1)  &--46.208(4)   &0.256(9)\\
   &                 &5  &2.99(3)  &--45.718(16)  &0.301(14)\\
   &                 &6  &1.46(3)  &--45.51(3)    &0.29(3)\\
   &                 &7  &0.71(1)  &--45.097(5)   &0.296(11)\\
\hline
2a &G111.526+0.803   &1  &15.73(3)  &--61.2860(3)    &0.367(1)\\
   &                 &2  &13.10(2)  &--60.6383(5)    &0.478(1)\\
   &                 &3  &0.59(2)   &--59.713(10)    &0.53(3)\\
   &                 &4  &3.30(12)  &--57.332(11)    &2.31(3)\\
   &                 &5  &17.49(10) &--58.0460(15)   &0.431(2)\\
   &                 &6  &16.05(11) &--57.5270(18)   &0.498(8)\\
   &                 &7  &6.3(2)    &--57.142(3)     &0.316(7)\\
   &                 &8  &7.44(9)   &--56.7108(15)   &0.396(5)\\
   &                 &9  &3.29(4)   &--55.866(2)     &0.423(6)\\
   &                 &10 &0.02(3)   &--52.9(2)       &0.2(5)\\
\hline
2b &G111.526+0.803   &1  &8.26(2)   &--61.3043(4)    &0.366(1)\\
   &                 &2  &7.00(2)   &--60.6560(6)    &0.474(2)\\
   &                 &3  &0.32(1)   &--59.730(12)    &0.51(3)\\
   &                 &4  &1.73(7)   &--57.338(13)    &2.28(4)\\
   &                 &5  &9.15(7)   &--58.0648(18)   &0.427(3)\\
   &                 &6  &8.56(7)   &--57.5427(19)   &0.513(10)\\
   &                 &7  &3.20(13)  &--57.155(3)     &0.304(8)\\
   &                 &8  &3.86(6)   &--56.7293(18)   &0.399(6)\\
   &                 &9  &1.71(2)   &--55.877(2)     &0.415(8)\\
\hline
3  &G111.532+0.759   &1  &38.66(6)  &--61.2903(3)   &0.366(1)\\
   &                 &2  &32.86(6)  &--60.6404(4)   &0.473(1)\\
   &                 &3  &1.45(6)   &--59.702(10)   &0.54(3)\\
   &                 &4  &8.0(3)    &--57.346(12)   &2.26(3)\\
   &                 &5  &43.8(3)   &--58.0500(15)  &0.427(2)\\
   &                 &6  &40.7(3)   &--57.5287(17)  &0.511(9)\\
   &                 &7  &15.4(6)   &--57.142(3)    &0.313(7)\\
   &                 &8  &18.4(2)   &--56.7134(15)  &0.400(5)\\
   &                 &9  &7.94(9)   &--55.863(2)    &0.419(6)\\
   &                 &10  &1.95(7)  &--52.995(6)    &0.339(13)\\
\hline
4  &G111.542+0.777   &1 &83.04(13)  &--61.3084(3)   &0.367(1)\\
   &                 &2 &70.90(12)  &--60.6578(4)   &0.473(1)\\
   &                 &3 &3.06(11)   &--59.719(10)   &0.53(2)\\
   &                 &4 &17.1(6)    &--57.373(11)   &2.25(3)\\
   &                 &5 &94.6(5)    &--58.0694(13)  &0.425(2)\\
   &                 &6 &88.0(6)    &--57.5455(15)  &0.519(8)\\
   &                 &7 &32.2(11)   &--57.157(2)    &0.308(6)\\
   &                 &8 &39.6(4)    &--56.7311(14)  &0.402(5)\\
   &                 &9 &17.10(18)  &--55.8796(19)  &0.415(6)\\
   &                 &10 &0.81(14)  &--53.01(3)     &0.32(6)\\
\hline
7  &G133.947+1.064   &1 &241(30)    &--45.514(6)   &0.282(17)\\
   &                 &2 &770(32)    &--45.494(8)   &0.635(9)\\
   &                 &3 &790(30)    &--45.080(2)   &0.296(6)\\
   &                 &4 &1784(4)    &--44.605(2)   &0.649(11)\\
   &                 &5 &778(25)    &--44.087(3)   &0.441(9)\\
   &                 &6 &1486(3)    &--43.569(1)   &0.512(4)\\
   &                 &7 &1169(19)   &--42.990(1)   &0.325(2)\\
   &                 &8 &545(20)    &--42.633(1)   &0.236(4)\\
   &                 &9 &125(11)    &--42.48(6)    &0.69(6)\\
   &                 &10 &13(2)     &--46.63(3)    &0.57(9)\\
\hline
10 &I 05137+3919   &1  &0.138(9)    &--20.704(10)  &0.32(2) \\ 
\hline 
11 &G173.482+2.446 &1  &13.29(2)   &--13.9420(2)   &0.246(1)\\
   &               &2  &16.96(3)   &--13.2253(4)   &0.271(1)\\
   &               &3  &3.52(2)    &--11.9252(7)   &0.267(2)\\
   &               &4  &5.28(13)   &--7.5164(7)    &0.229(3)\\
   &               &5  &4.21(1)    &--13.587(4)    &1.121(6)\\
   &               &6  &1.48(2)    &--11.775(4)    &0.849(8)\\
   &               &7  &2.63(13)   &--7.467(3)    &0.420(6)\\
   &               &8  &1.84(4)    &--12.935(3)    &0.272(7)\\
\hline
13 &IRAS 05382+3547 &1 &0.038(5)   &--24.201(12)   &0.18(3)\\
   &                &2 &0.034(4)   &--23.781(16)   &0.28(4)\\
\hline
15 &G183.349--0.575 &1 &13.16(3)   &--15.2058(3)  &0.2746(7)\\
   &                &2 &6.66(2)    &--14.5972(7)  &0.3734(17)\\
   &                &3 &0.46(2)    &--13.990(9)   &0.34(2)\\
   &                &4 &0.70(3)    &--5.118(19)   &0.71(3)\\
   &                &5 &9.25(5)    &--4.8540(5)   &0.2543(15)\\
\hline
16a &G213.705--12.60 &1 &29.31(10)  &10.5536(5)  &0.3726(15)\\
    &               &2 &13.21(13)  &10.9728(9)  &0.248(3)\\
    &               &3 &68.50(7)   &12.9606(3)  &0.4281(9)\\
    &               &4 &18.74(7)   &13.4374(10) &0.342(2)\\
    &               &5 &2.42(13)   &12.373(5)   &0.313(17)\\
    &               &6 &0.99(7)    &9.992(11)   &0.31(3)\\
    &               &7 &0.67(5)    &11.63(5)    &1.1(3)\\
\hline 
16b &G213.705--12.60 &1 &25.98(10)  &10.5514(5)   &0.3686(16)\\
    &               &2 &11.68(13)  &10.9701(10)  &0.252(3)\\
    &               &3 &66.64(7)   &12.9651(3)   &0.4281(9)\\
    &               &4 &17.49(7)   &13.4402(11)  &0.344(2)\\
    &               &5 &2.25(12)   &12.381(5)    &0.323(18)\\
    &               &6 &0.90(6)    &9.9931(12)   &0.319(29)\\
    &               &7 &0.62(4)    &11.63(6)     &1.2(3)\\
\hline
17 &G189.030+0.783  &1 &10.7(2)   &8.9173(5)    &0.251(1)\\
   &                &2 &2.14(11)  &9.101(14)    &0.377(13)\\
   &                &3 &4.22(17)  &9.6742(13)   &0.189(2)\\
   &                &4 &5.21(8)   &9.824(3)     &0.274(3)\\
\hline
18 &G188.946+0.886  &1 &25.2(5)   &9.6476(8)     &0.279(4)\\
   &                &2 &27(2)     &10.32(6)      &0.77(8)\\
   &                &3 &250(5)    &10.4921(12)   &0.291(2)\\
   &                &4 &234(3)    &10.6867(6)    &0.1987(13)\\
   &                &5 &394(39)   &10.8800(7)    &0.241(4)\\
   &                &6 &286(29)   &10.987(13)    &0.319(8)\\
   &                &7 &7.96(12)  &11.542(2)     &0.231(5)\\
   &                &8 &3.36(17)  &8.475(9)      &0.288(18)\\
   &                &9 &1.57(9)   &8.85(3)       &0.42(6)\\
   &               &10 &0.31(7)   &7.86(7)       &0.53(17)\\
\hline
19 &G196.454--1.677 &1 &0.33(1)   &13.855(6)     &0.190(13)\\
   &                &2 &0.28(1)   &14.126(8)     &0.24(2)\\
   &                &3 &6.3(3)    &14.5855(8)    &0.175(3)\\
   &                &4 &14.69(10) &14.741(2)     &0.288(3)\\
   &                &5 &8.37(2)   &15.1184(5)    &0.233(2)\\
   &                &6 &1.45(8)   &15.2951(14)   &0.121(4)\\
   &                &7 &7.78(2)   &15.4524(9)    &0.283(2)\\
   &                &8 &3.12(1)   &15.8930(5)    &0.234(1)\\
\hline
20 &G232.620+0.996  &1 &8.5(2)    &21.619(5)     &0.234(11)\\
   &                &2 &17.8(2)   &21.949(3)     &0.250(8)\\
   &                &3 &59.0(2)   &22.3075(9)    &0.303(2)\\
   &                &4 &198.7(2)  &22.9306(2)    &0.3377(4)\\
   &                &5 &3.2(2)    &23.702(9)     &0.24(2)\\
\hline
21a &G240.316+0.071 &1 &0.054(6)  &62.84(2)      &0.22(5)\\
    &               &2 &0.234(9)  &63.208(5)     &0.256(13)\\
\hline 
21b &G240.316+0.071 &1 &0.040(8)  &62.85(6)      &0.39(15)\\
    &               &2 &0.371(12) &63.211(5)     &0.237(10)\\ 
\hline
\end{longtable*}

\end{document}